\documentclass[11pt]{article}

\usepackage{multirow}
\usepackage{graphics}
\usepackage{graphicx}
\usepackage{color,xcolor}
\usepackage{ulem}

\usepackage{amsmath,amssymb,amsthm}

\usepackage{lineno,hyperref}
\modulolinenumbers[5]

\usepackage{natbib}

\title{Point-process based Bayesian modeling of space-time structures of forest fire occurrences in Mediterranean France}

\author{Thomas Opitz$^a$, Florent Bonneu$^b$, Edith Gabriel$^{ab}$, \\
	\small{$^a$Biostatistics and Spatial Processes, INRAE, 84914 Avignon, France}\\
	\small{$^b$LMA EA2151, Avignon University, 84000 Avignon, France}
}

\begin{document}
\maketitle 





\begin{center}
	\begin{minipage}{0.8\linewidth}
{\textbf{Abstract. }}
Due to climate change and human activity, wildfires are expected to become more frequent and extreme worldwide, causing  economic and ecological disasters. The deployment of preventive measures and operational forecasts can be aided by stochastic modeling that helps to understand and quantify the mechanisms governing the occurrence intensity. We here develop a point process framework for wildfire ignition points observed in the French Mediterranean basin since 1995, and we fit a spatio-temporal log-Gaussian Cox process with monthly temporal resolution  in a Bayesian framework using the integrated nested Laplace approximation (INLA).
Human activity is the main direct cause of wildfires and is indirectly measured through a number of appropriately defined proxies related to land-use covariates (urbanization, road network) in our approach, and we further integrate covariates of climatic and environmental conditions to explain wildfire occurrences.  We include spatial random effects with Mat\'ern covariance and temporal autoregression at yearly resolution. Two major methodological challenges are tackled : first, handling and unifying multi-scale structures in data is achieved through computer-intensive preprocessing steps with GIS software and kriging techniques; second, INLA-based estimation with high-dimensional response vectors and latent models is facilitated through  intra-year subsampling, taking into account the occurrence structure of wildfires. 

\end{minipage}
\end{center}



\section{Introduction}
In the French territories close to the Mediterranean, the number of wildfire occurrences and their burnt surfaces have reached alarmingly high levels due to evolutions such as the increase in human activity, often at the interface with forest areas, and climate change. In $2017$ alone, $2,300$ forest fire occurrences totalling $20,000$ hectares of burnt surface have been reported. For efficient risk prevention, we  have to understand the stochastic mechanisms governing the spatial and temporal variability in the intensity of wildfire occurrences. In this context, statistical modeling of forest fires is crucial to identify the main drivers of fire occurrences and to develop operational forecasting tools. With the point-process based modeling of space-time structures in  wildfire occurrences proposed in this work, we take a major step in this direction by taking into account climatic and environmental trends and complex residual interaction in the spatio-temporal distribution.

Wildfires have already been studied with point process approaches to assess how the spatial heterogeneity of wildfires observed over a given time interval depends on the spatial distribution of land use information such as vegetation, urban zones or wetlands \citep{juan2012, moller2010, pereira2013, serra2014}. In practice, raw data of environmental covariates is usually only available at very specific spatial and temporal scales, often with very high resolution, and comes in different numerical formats. Appropriate preprocessing of such data is important to obtain good predictive models. Raw data may also have a low signal-to-noise ratio with respect to forest fire prediction,  and one can obtain better predictive relevance by generating artificial covariates that summarize relevant conditions. While such preprocessing steps are often not implemented  or explained  in the literature, we will here put focus on how we preprocess raw data to obtain meaningful covariates for driving the point process intensity. 

The high spatio-temporal dimension of wildfire data (observed occurrences and control cases without occurrences) has often been coped with by separating the data into subsets or by strongly  aggregating them, by year or by spatial areas \citep[e.g.,][]{genton2006, serra2014, turner2009, xu2011}. Some recent approaches have concentrated more strongly  on studying the interplay of the spatial and temporal structures \citep{Gabriel2017}, or on the usefulness of a specific Fire Weather Index aggregating weather data \citep{Fargeon.al.2018}. In the following, we choose a monthly temporal resolution to be able to capture the intra-year seasonality of weather influences and other effects. 

We here use log-Gaussian Cox process models, which have already been identified as useful models for wildfires since they allow capturing  spatio-temporal aggregation structures through random effects  \citep{Gabriel2017,pereira2013,serra2014}. 
Since we consider a monthly time scale over $24$ years with a spatial resolution given by a $2km$-grid for wildfire occurrence records and covariates, established by the the French authority responsible for forest fires (``D\'{e}fense de la for\^{e}t contre les incendies", DFCI in short), very high-dimensional response vectors and covariate matrices arise in our models. Bayesian inference for log-Gaussian Cox processes using the integrated nested Laplace approximation \citep[INLA, see][]{Rue.al.2009,Illian.al.2012} is now well-established, but remains challenging with the high dimension of our regression model. To overcome this difficulty, we here develop a stratified subsampling technique to reduce the length of the covariate vector. This method is designed to "trade space for time" and relies on the strong small-scale spatial dependence in climatic covariates and seasonal behavior; at each DFCI grid cell it subsamples the months of the year where no fire occurred. This technique allows us to divide  approximately by $11$ the size of the response vector used with INLA. 

The remainder of the paper is organized as follows. Section~\ref{sec:data} describes data, related to forest fire occurrences, environmental and climatic covariates. It details the GIS-based preprocessing steps, and for weather variables we use monthly station data and interpolate them by spatio-temporal kriging. Section~\ref{sec:StLGCP} develops our log-Gaussian Cox process models with different spatial-temporal structures for the random effects.  Section~\ref{sec:bayes} is dedicated to the Bayesian estimation method where we detail our choices for prior distributions with INLA, and we present the subsampling technique of intra-year effects. We report and interpret parameter estimation results and intensity prediction of our most complex model in Section~\ref{sec:res}. Finally, we conclude with a  discussion of our approach and further perspectives in Section~\ref{sec:conc}.

\section{Data on fire occurrences and covariates}
\label{sec:data}
\subsection{Prom\'eth\'ee fire occurrence dataset}

\begin{figure}
\centering
\includegraphics[height=8cm]{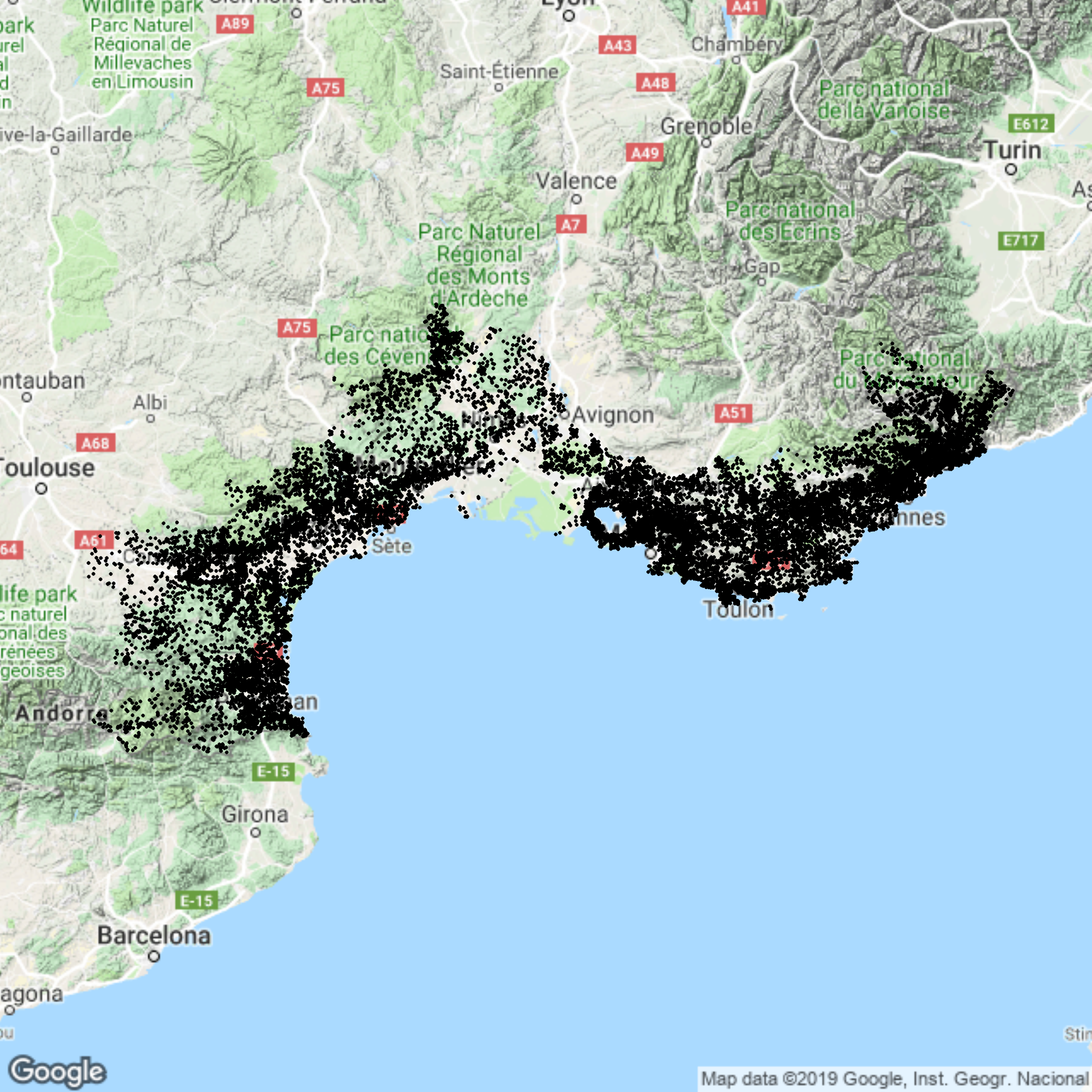}
\caption{Map showing the $23,309$ wildfire locations observed during the $1995-2018$ period over the $9,562$ DFCI grid cells in the Mediterranean basin, Southern France.}
\label{fig:data13}
\end{figure}

Since 1973, the French government maintains the \emph{Prom\'eth\'ee}  database of forest fire occurrences in Southern France to allow for the  development of  statistical tools to gain better knowledge of the spatial and temporal distribution of wildfires and their causes. We consider a subset of this database by selecting all wildfire occurrences, with burnt areas more than 1 hectare, in the years from 1995 to 2018 in the French Mediterranean basin, which is composed of seven ``d\'{e}partements"\footnote{Pyr\'en\'ees-orientales, Aude, H\'erault, Gard, Bouches-du-Rh\^one, Var, Alpes-maritimes.} with overall surface of about $40,000$ $km^2$. 
We focus on the years 1995 to 2018 since they correspond to the time period where our purely spatial environmental  covariate dataset has been established. The spatial resolution of wildfire reports is given by the DFCI coordinates spanning a grid in the Lambert$93$ projection with quadratic cells covering approximatively $4km^2$ each. The study area is  partitioned into $9,562$ DFCI cells. Therefore, there is some small-scale positional uncertainty since we do not know the exact locations of wildfire occurrences inside the cells. 
Figure~\ref{fig:data13} shows a map of the $23,309$ wildfire occurrence positions observed during the study period. 

\subsection{Land use and land cover}

\begin{figure}
\centering
\includegraphics[width=0.49\linewidth]{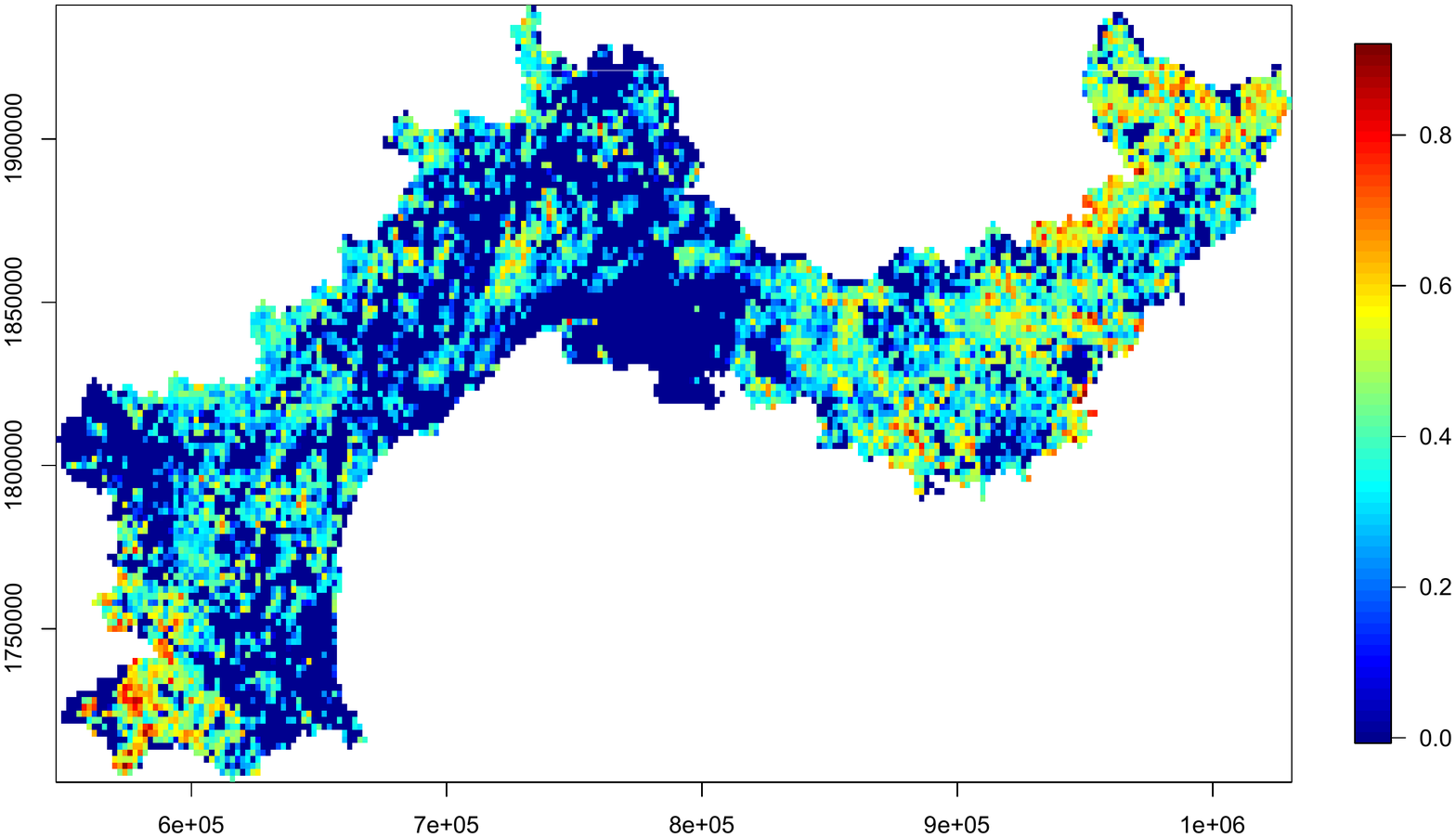}~
\includegraphics[width=0.49\linewidth]{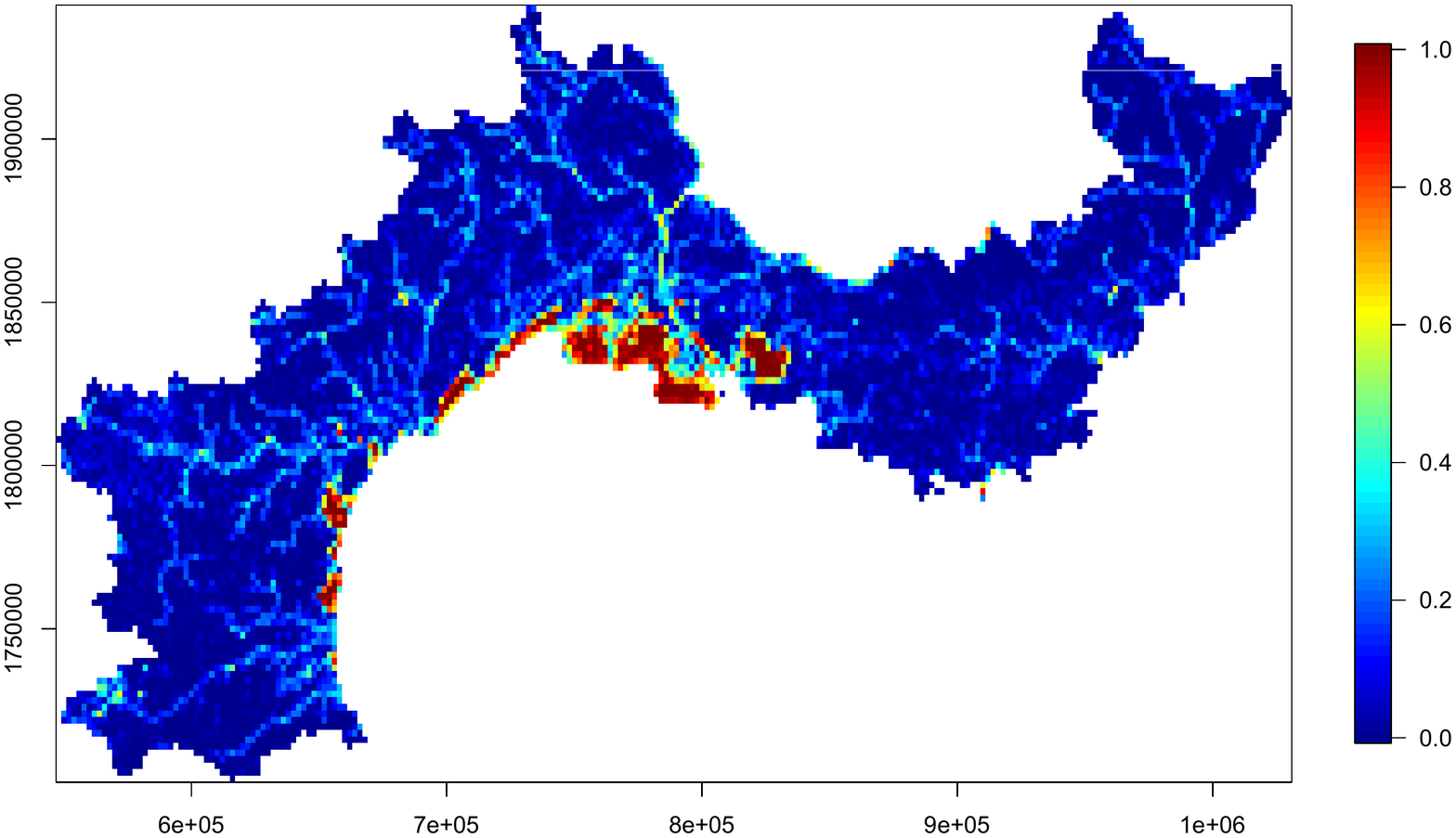}

\includegraphics[width=0.49\linewidth]{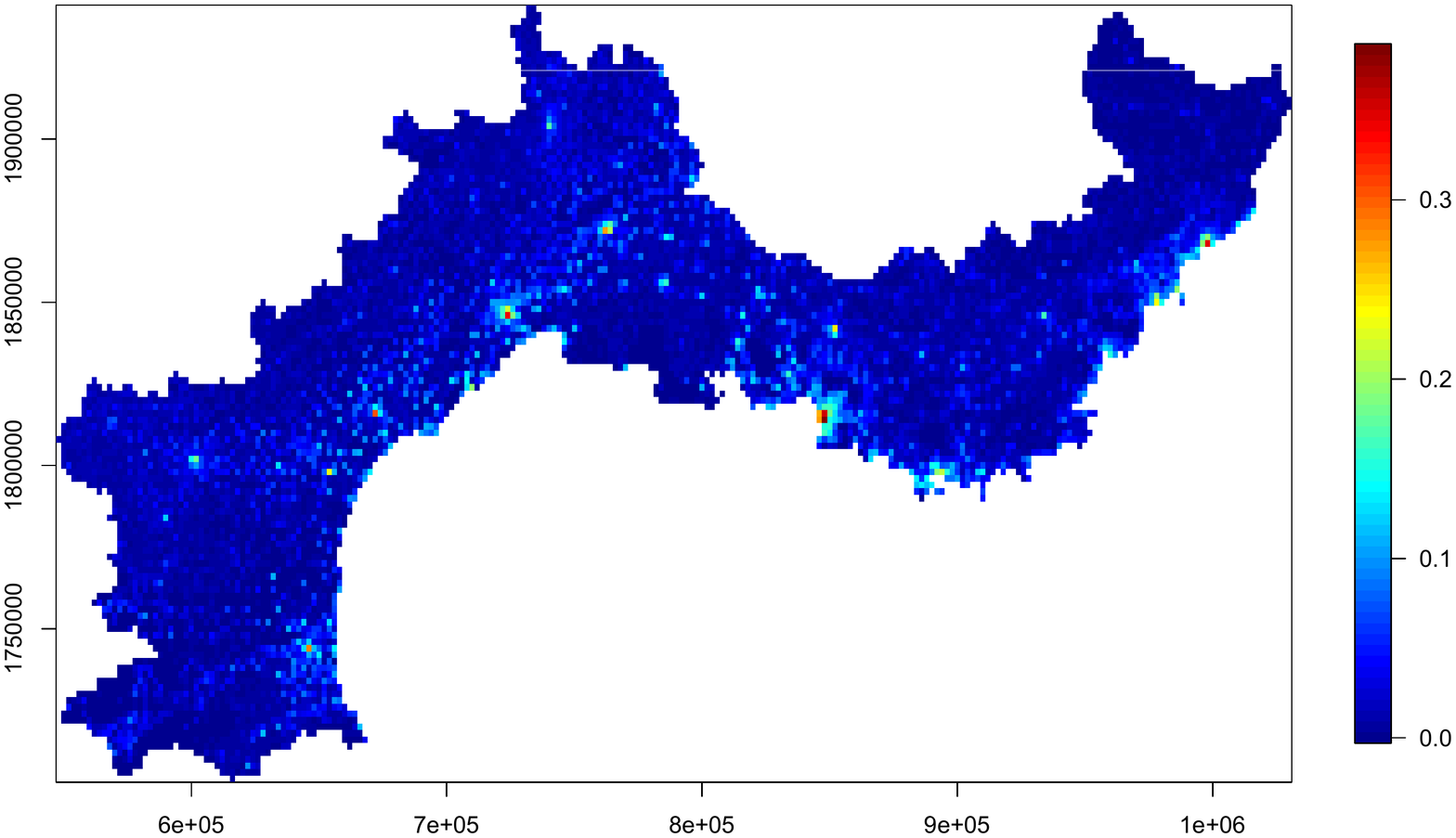}~
\includegraphics[width=0.49\linewidth]{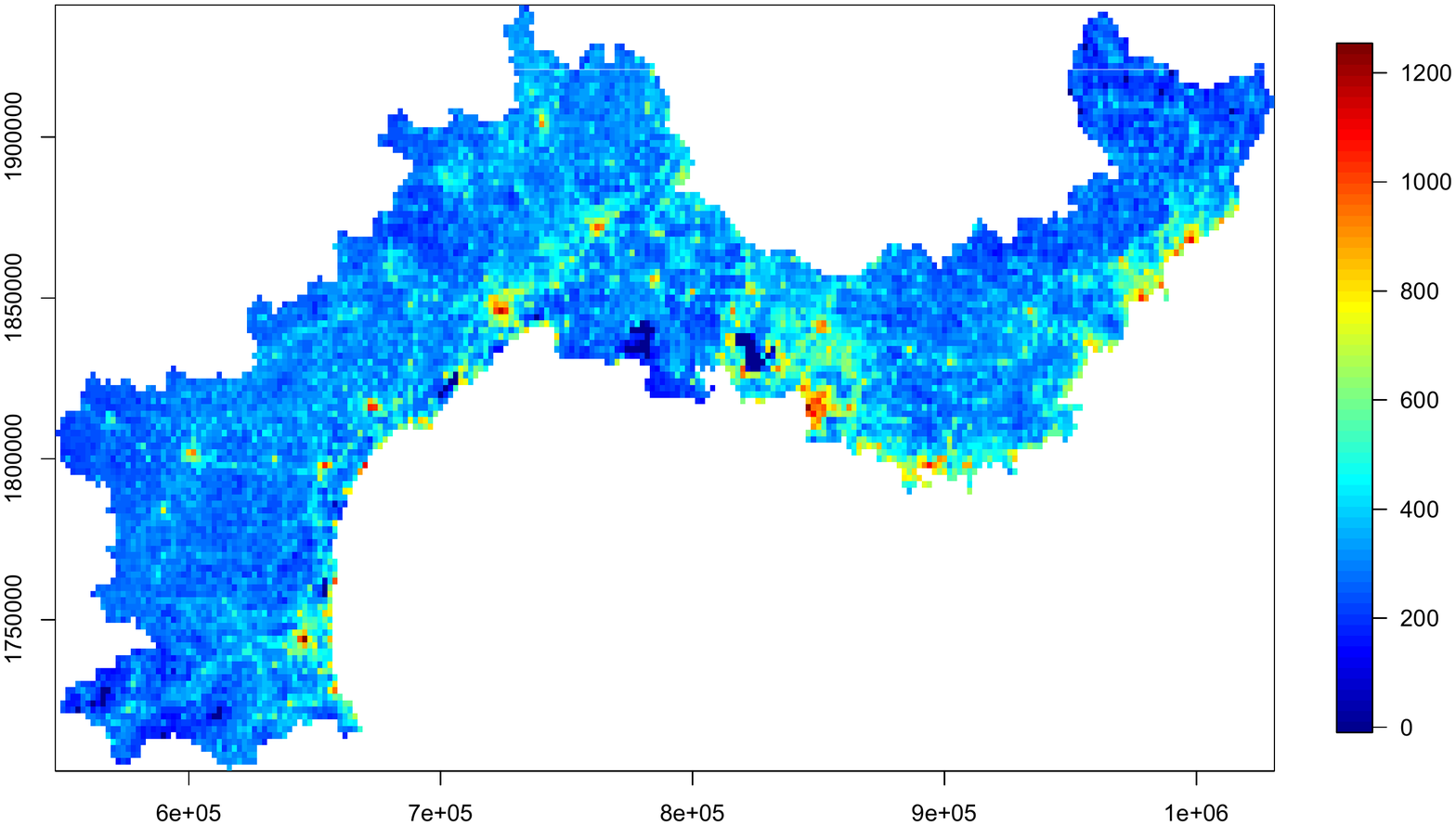}
\caption{Examples of land cover covariates. Top left: coverage of coniferous trees; top right: water coverage; bottom left: building coverage; bottom right: total length of roads (in kilometers).}
\label{fig:covariates}
\end{figure}

Land use and land cover information is very useful to model the probability of wildfire occurrences. The concomitance of environmental conditions (water, vegetation type…) and human infrastructure (buildings, road networks) is often crucial to explain wildfire outbreaks, which occur most of the time due to human negligence. We exploit several land use databases\footnote{BD FORET V1, BD TOPO and BD ALTI}, freely provided by the French National Institute of Geographic and forest information (IGN) for research purposes, to generate land use covariates. IGN maps are available for each ``d\'epartement" in several formats (shapefiles or .tiff images) with different spatial resolution (2m, 25m…) and geometry types (pixels, lines, polygons). We extracted $14$ covariates, subdivided into 4 groups: 1) water, elevation and slope; 2) protected zones with special regulatory rules for tourism; 3) vegetation type: open forest, coniferous forest, shrubland, moorland, conifers and coppices; 4) urbanization: building cover, primary roads, secondary roads, roads and paths. Maps of several of the land cover covariates are shown in Figure~\ref{fig:covariates}.

We first unify these datasets with different spatial resolution towards the DFCI grid. By computer-intensive preprocessing steps, raw datasets are transformed to summary covariates of land cover on the DFCI grid. For each covariate, we provide a single contiguous spatial map by using GIS software (QGIS) to aggregate the databases available for the seven ``d\'epartements".  Next, we use the recent R package \texttt{sf}, allowing us to handle spatial geometric objects, to compute the proportion of millions of polygon areas of forests and buildings, and the overall road lengths for the cells of an intermediate regular grid with $200$ meter resolution. Finally, for each DFCI cell at $2km$ resolution and each covariate, we compute the mean and standard deviation of the $100$ values in the $200$m-subgrid inside each DFCI cell. This two-step approach has the advantage to keep valuable information concerning the fine-scale structure of covariate values when we aggregate them to the DFCI grid. The mean summary shows the overall trend inside each DFCI cell, and the standard deviation measures the  variability around this trend and determines small-scale heterogeneity.  

Another important objective  is to generate synthetic covariates that highlight the interface of forested areas to human activity. Indeed, wildland-to-urban interfaces are heterogeneous areas where the two main drivers of wildfires,  vegetation and human activity, are expected to strongly coincide with outbreaks. We compute two additional covariates to take into account the interface between open forests and urbanized areas, and between open forests and paths. For this purpose, we multiply the percentage of open forests by respectively the percentage of buildings and the total length of paths inside each DFCI cell. 

Land use is a dynamic process evolving over time, and some minor to moderate changes may have taken place during the study period, whereas we have only static spatial data for the covariates described above. Indeed, the census process used by IGN is long and spans several years to provide complete maps for regions such as the French Mediterranean area. We point out that the  impact of environmental changes may be relatively minor in our case since the study period corresponds to the census period for the IGN databases that we use here; they provide a  summary of partial measurements  over the time period $1995-2000$. 

\subsection{Monthly weather data}
\begin{figure}
\centering
\includegraphics[width=0.49\linewidth]{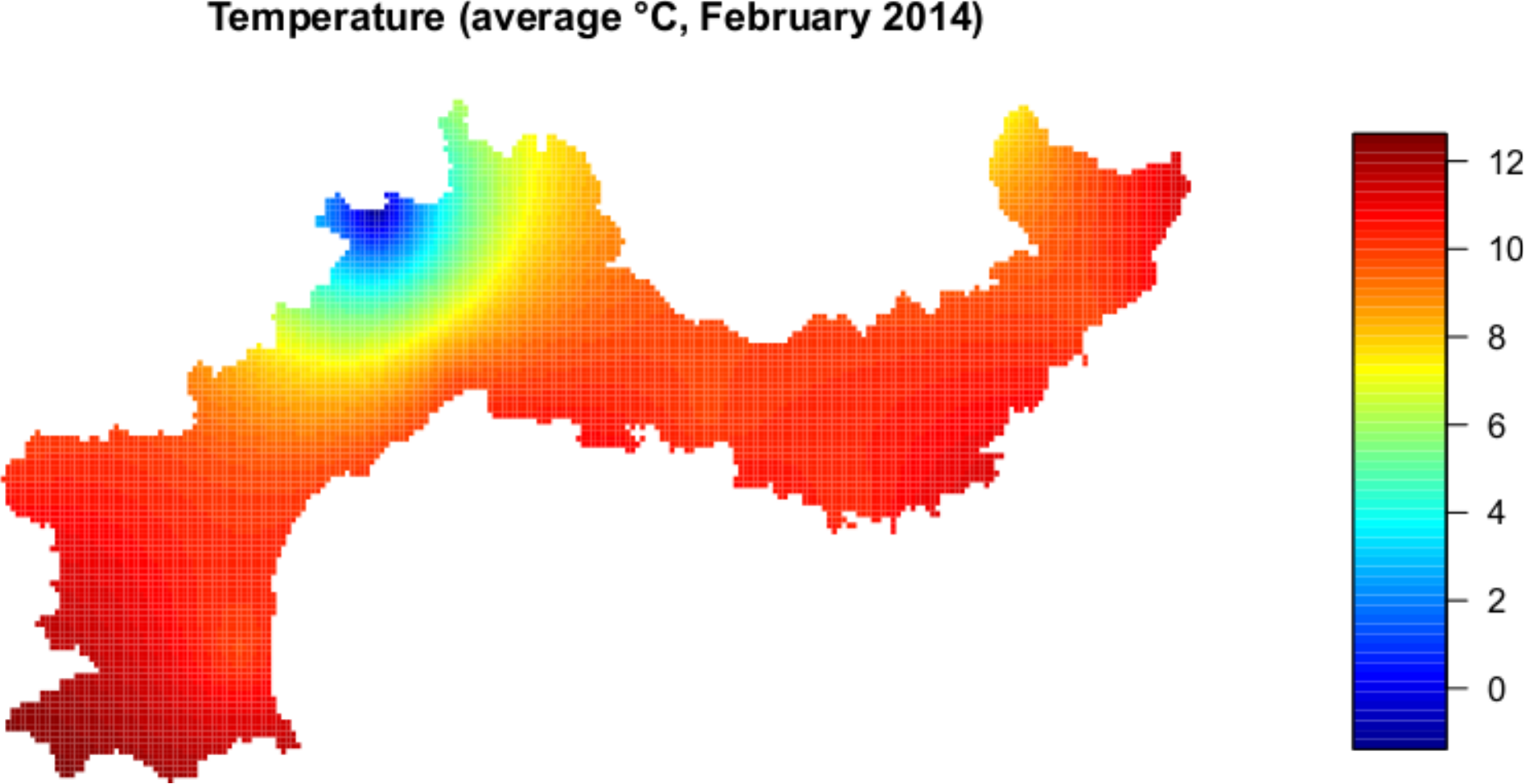}~
\includegraphics[width=0.49\linewidth]{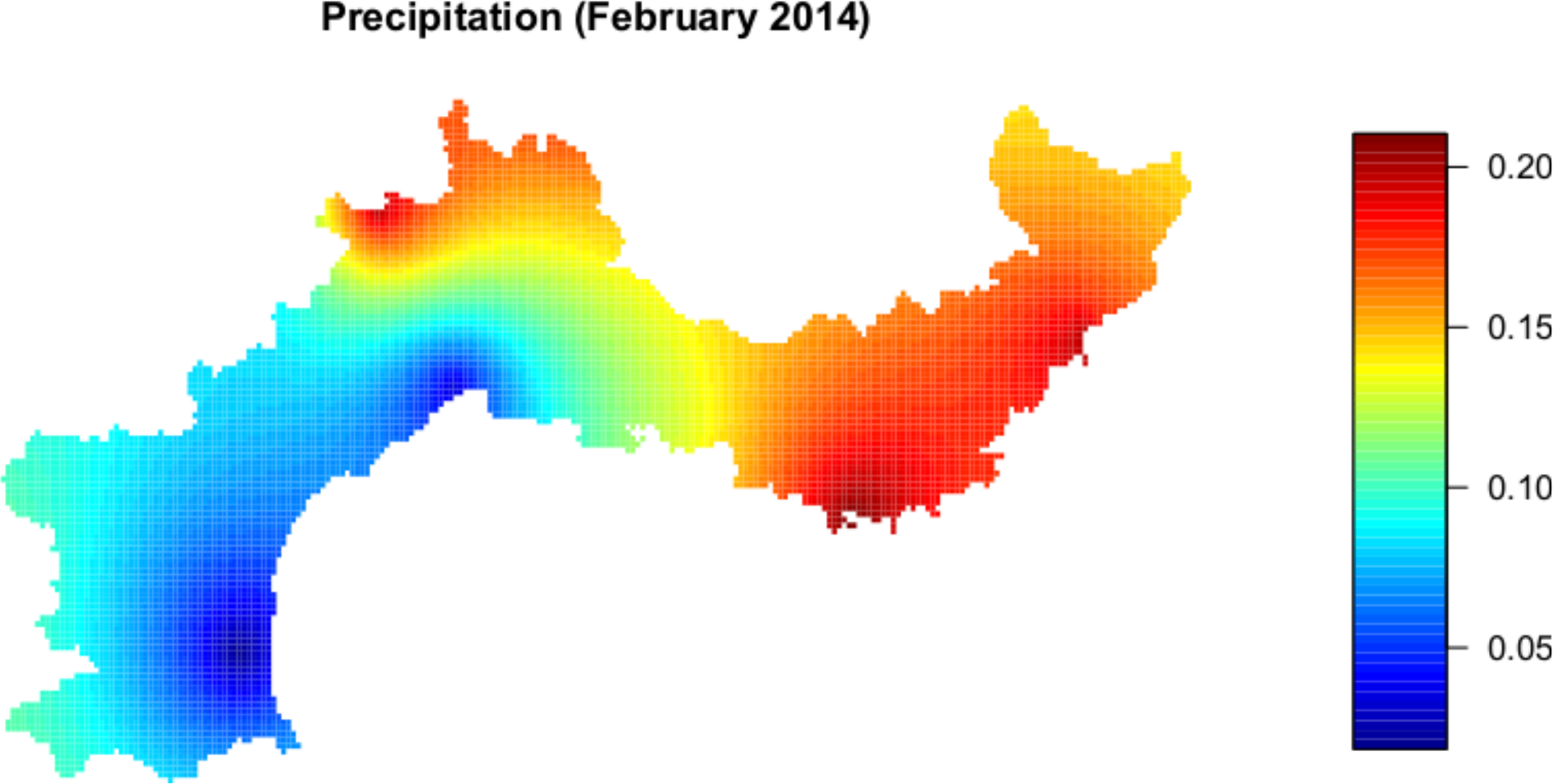}

\includegraphics[width=0.49\linewidth]{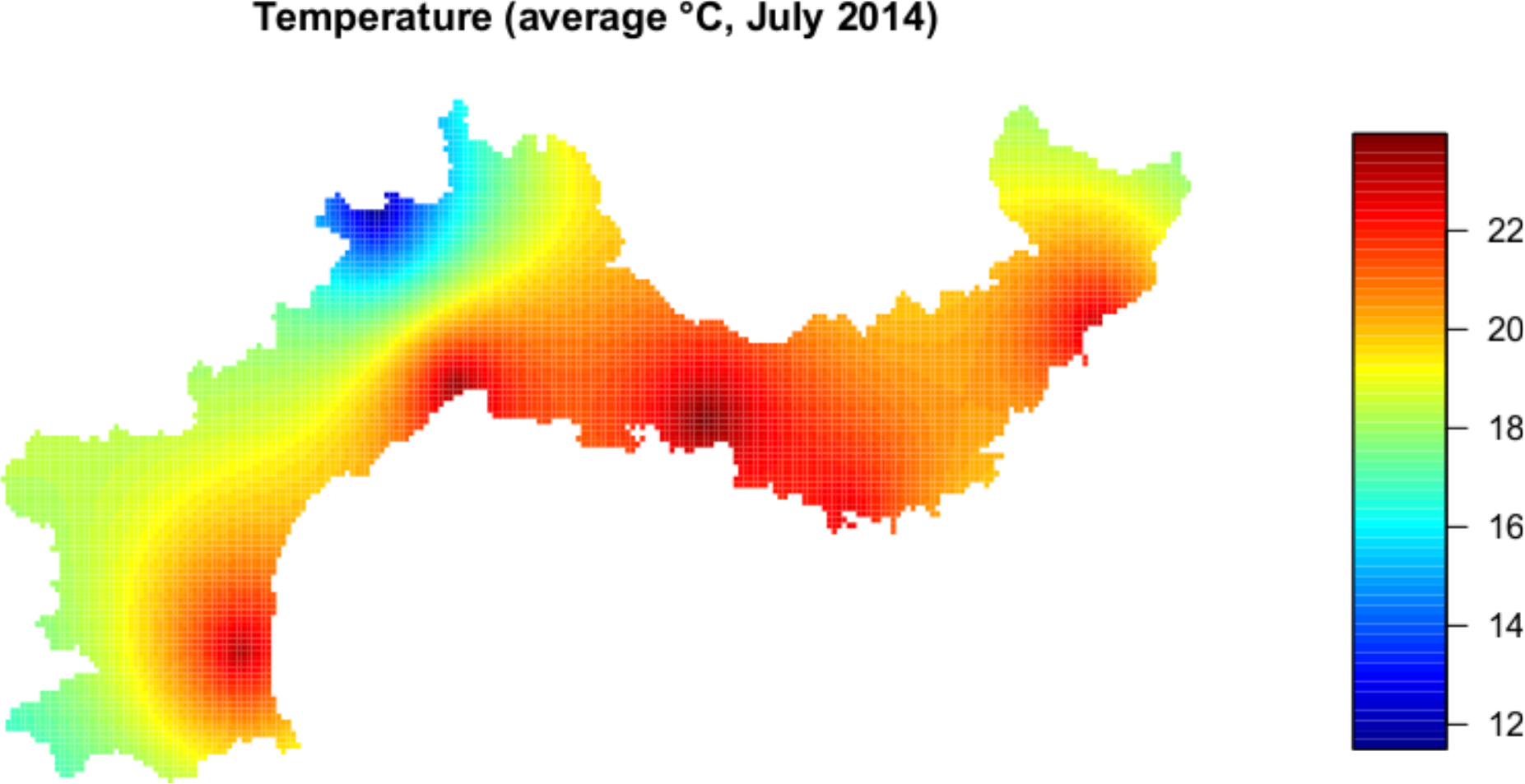}~
\includegraphics[width=0.49\linewidth]{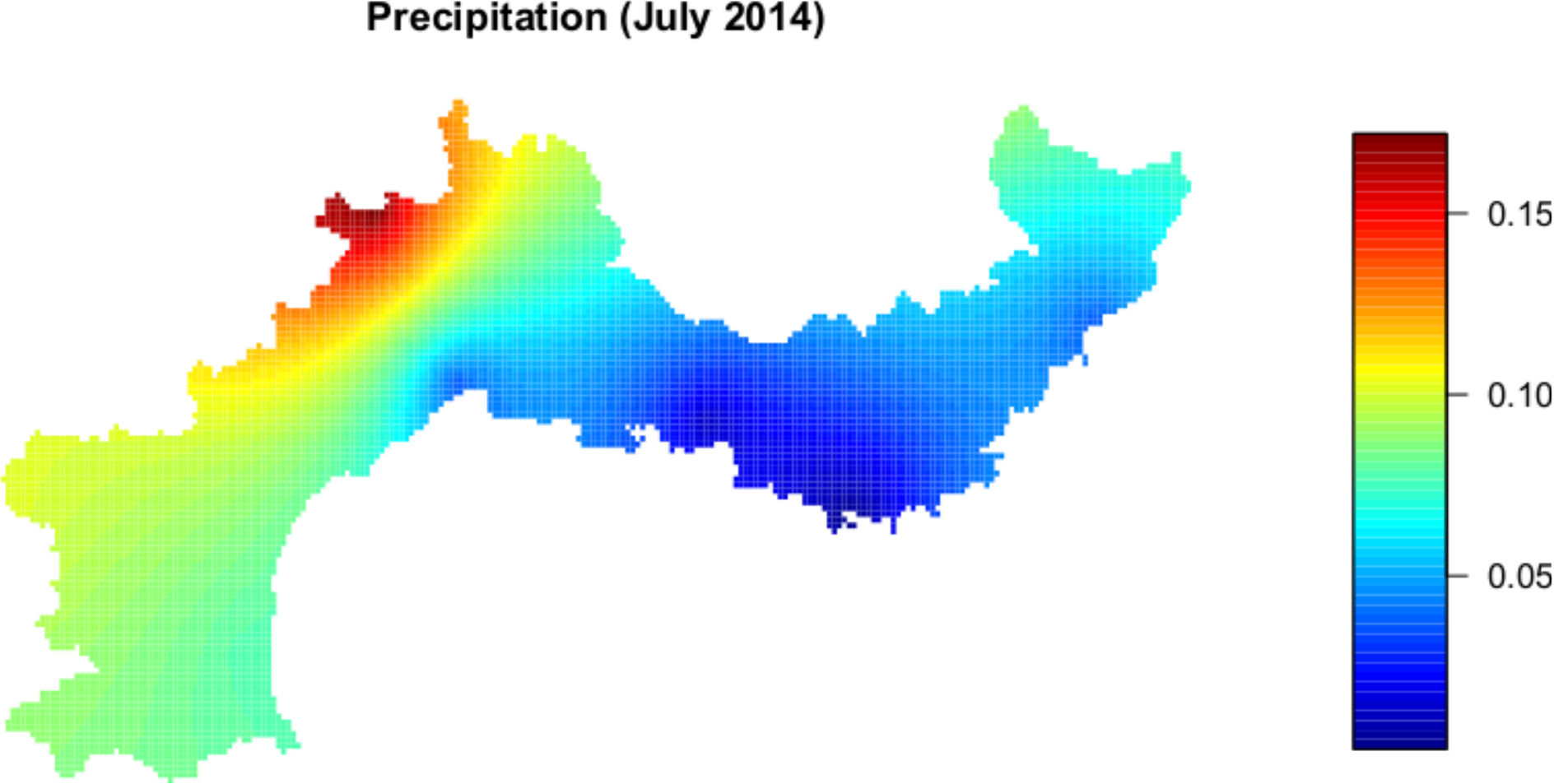} 
\caption{Examples of monthly weather covariates obtained by kriging interpolation. Left: temperature; right: precipitation.}
\label{fig:krigmeteo}
\end{figure}

Weather plays an important role for forest fire outbreaks. Often, favorable conditions such as high temperature and low precipitation leading to very dry soil and vegetation build up over several months and increase the probability of wildfire  occurrences. We here consider monthly temperature and precipitation height data (averages of daily data) obtained from the National Oceanic and Atmospheric Administration (NOAA), with units Celsius and inches, respectively.  These data have been recorded at $17$ weather stations in the French Mediterranean area and contain the minimum, maximum and average temperature, and the total precipitation,  denoted by TMIN, TMAX, TAVG and PRCP, respectively, and we first transform these data to monthly averages at each station. To spatially interpolate the local weather conditions for each DFCI cell, we then apply spatio-temporal kriging. After some preliminary analyses using the {\tt gstat} \texttt{R} package \cite{Pebesma2004}, we decide to implement product-sum variograms of \cite{Deiaco2001} for the temperature and a space-time separable variogram for the square-root of the precipitation, all with exponential submodels for the spatial and temporal components. 
The product-sum variogram model is defined as
\begin{equation}
\label{eq:vario_productsum}
    \gamma (h,u) = \left(k\cdot \text{sill}_t + 1 \right) \gamma_s (h) + \left(k\cdot \text{sill}_s + 1 \right) \gamma_t (u) - k \gamma_s(h) \gamma_t(u),
\end{equation}
where $\gamma_s$ and $\gamma_t$ are spatial and temporal variograms (with spatial and temporal sills $\text{sill}_s$ and $\text{sill}_t$, respectively), and $k$ is a positive parameter. 
The separable variogram model is
\begin{equation}
\label{eq:vario_separable}
    \gamma (h,u) = \text{sill} \times  \left( \bar \gamma_s (h) + \bar \gamma_t (u) -  \bar \gamma_s(h) \bar \gamma_t(u) \right),
\end{equation}
where $\bar \gamma_s$ and $\bar \gamma_t$  are standardized spatial and temporal variograms with separate nugget effects and (joint) sill $1$. The overall sill parameter is denoted by ``$\text{sill}$''.
All models are fitted using {\tt gstat}. Table~\ref{tab:variog} provides the estimated parameters of models (\ref{eq:vario_productsum}) and (\ref{eq:vario_separable}) for the four weather variables. 
\begin{table}
	\centering 
\begin{tabular}{|l|c|c|c|c|c|}
     \hline 
     & \multicolumn{5}{|c|}{Product-sum model with exponential components} \\
     \hline
     & \multicolumn{2}{|c|}{Spatial component} &  \multicolumn{2}{|c|}{Temporal component} & \\
     Variable & $sill_s$ & range & $sill_t$ & range & $k$ \\
     \hline
     TAVG &  46.29 & 60km & 99.98 & 3.97 & $1.49\times   10^-8$\\
    TMIN & 46.28 & 60km &  149.95 & 11.73 & $1.49 \times  10^-8$ \\
    TMAX & 61.72 & 60km &  61.72 &  8.75 & $1.49 \times 10^-8$ \\
     \hline
           \hline & \multicolumn{5}{|c|}{Separable model with exponential component} \\
     \hline
     & \multicolumn{2}{|c|}{Spatial component} &  \multicolumn{2}{|c|}{Temporal component} & \\
     Variable &  (nugget, $sill_s$) & range& (nugget, $sill_t$) & range & $sill$ \\
     \hline
     PRCP &(0.562, 0.438) & 60km & (0.562, 0.438) & 5.33 & 0.019 \\
     \hline
\end{tabular}
\caption{Estimated spatio-temporal variogram models. The temporal unit is $1$ month.}
\label{tab:variog}
\end{table}
For illustration, Figure~\ref{fig:krigmeteo} shows interpolated maximum temperature (Celsius) and precipitation in inches over the French Mediterranean region for the months of February and July 2014. 

\begin{figure}
\centering
\includegraphics[width=0.48\linewidth]{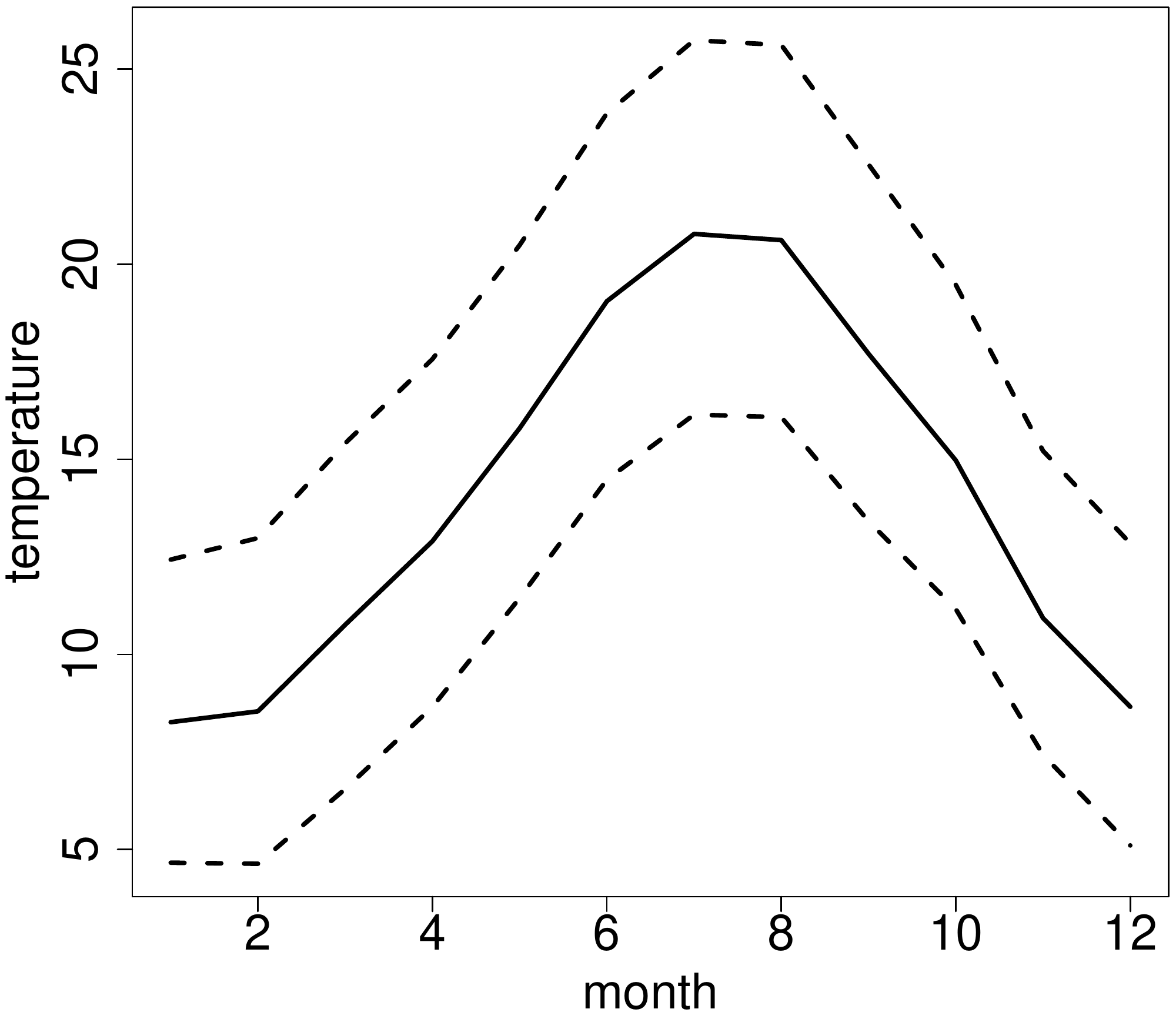}$\quad$ 
\includegraphics[width=0.48\linewidth]{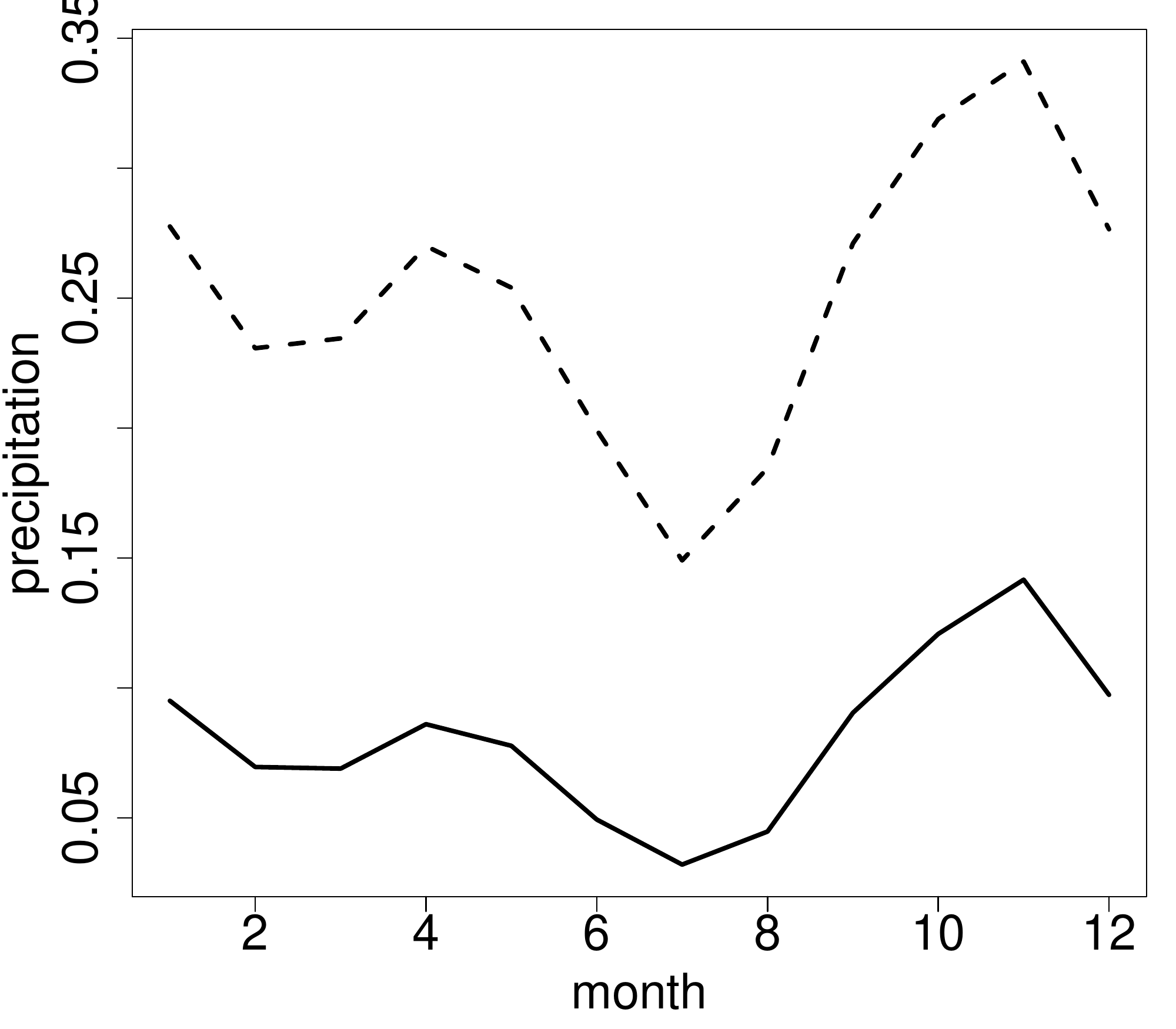} 
\caption{Seasonal (monthly) trends in spatially interpolated weather data. Left: TAVG (solid curve), and TMIN, TMAX (dashed curves). Right: precipitation (solid curve); square root of precipitation (dashed curve).}
\label{fig:trendclimate}
\end{figure}

In the construction of our models in Section~\ref{sec:StLGCP}, we aim to  separate a  seasonal effect of fire intensity (expressed on a monthly scale) from weather-related effects. To better decorrelate the weather observations from seasonal behavior and to focus on the influence of the weather anomalies with respect to typical climatic conditions, we also  calculate the averages for each of the 12 months over the full study area and period, and we then subtract these monthly averages from the monthly kriging interpolations to obtain our final weather (anomaly) covariates. The estimated global average monthly effects, used for calculating monthly weather anomalies, are shown in Figure~\ref{fig:trendclimate}. Further analyses have shown that the three monthly temperature variables TAVG, TMIN and TMAX are highly correlated, with linear correlation coefficients larger than  $0.99$ for kriging predictions and larger than $0.97$ for anomalies. Therefore, we avoid identifiability issues in our model by keeping only the anomalies of TAVG as a covariate in the models proposed in Section~\ref{sec:StLGCP}. Moreover, to appropriately capture the influence of precipitation on wildfire occurrences, we use two covariates: anomalies of monthly precipitation, and anomalies in the square root of monthly precipitation. The latter has different structure, characterized by higher amplitudes, and the combination of both variables allows us to capture certain nonlinearities in the precipitation effect, even if we estimate only linear regression coefficients in our models.

\section{Spatio-temporal log-Gaussian Cox process models}
\label{sec:StLGCP}

Conceptually, we can think of our model as defined over continuous space and time. In practice, we here choose a discretization of space (DFCI grid cells) and of time (months), such that the random intensity $\Lambda(s,t)$ does not vary within one DFCI cell during one month when estimating the model. The center of the DFCI grid cell is used as a representative coordinate when working with random spatial effects. 

\subsection*{Model structure}
The log-intensity of the point process in our model has the following additive structure, including $30$ covariates $z_j^{\mathrm{land}}$ related to land use and $3$ covariates $\hat{z}_j^{\mathrm{clim}}$ related to weather conditions: 
\begin{align}\label{eq:Lambda}
\log \Lambda(s,t)=&\beta_0+\beta^{\mathrm{time}}\tilde{t}+\sum_{j=1}^{30} \beta_j^{\mathrm{land}} z_j^{\mathrm{land}}(s)+\sum_{j=1}^{3} \beta_j^{\mathrm{clim}} \hat{z}_j^{\mathrm{clim}}(s,t)\\
&+ f(\mathrm{month}(t))+W(s,a(t))
\end{align}
where the first line shows fixed effects with coefficients $\beta_j^\mathrm{type}$ to estimate, and the second line shows random effects. A Gaussian space-time random effect $W(s,a(t))$ is defined at the level of the year $a(t)$ associated with the month $t$ (overall $282$ months), i.e. $a(t)\in \{1995,\ldots,2018\}$. Moreover, based on a transformation $\tilde{t}=\frac{t-t_{\min}}{t_{\max}-t_{\min}}$  of the observation period to the interval $[0,1]$, a seasonal effect $f(\mathrm{month}(t))$ is 
 defined at monthly resolution with $12$ levels. 
We here use the hat-notation $\hat{z}_j^{\mathrm{clim}}$ to underline that climate covariates have been estimated beforehand through kriging of observations on an irregular grid of $17$ weather stations. Therefore, the influence of the kriging uncertainty on landslide intensity predictions could be studied in practice, although we will not pursue this idea here but discuss this point in Section~\ref{sec:conc}.

\begin{figure}
    \centering
    \includegraphics[width=0.475\linewidth]{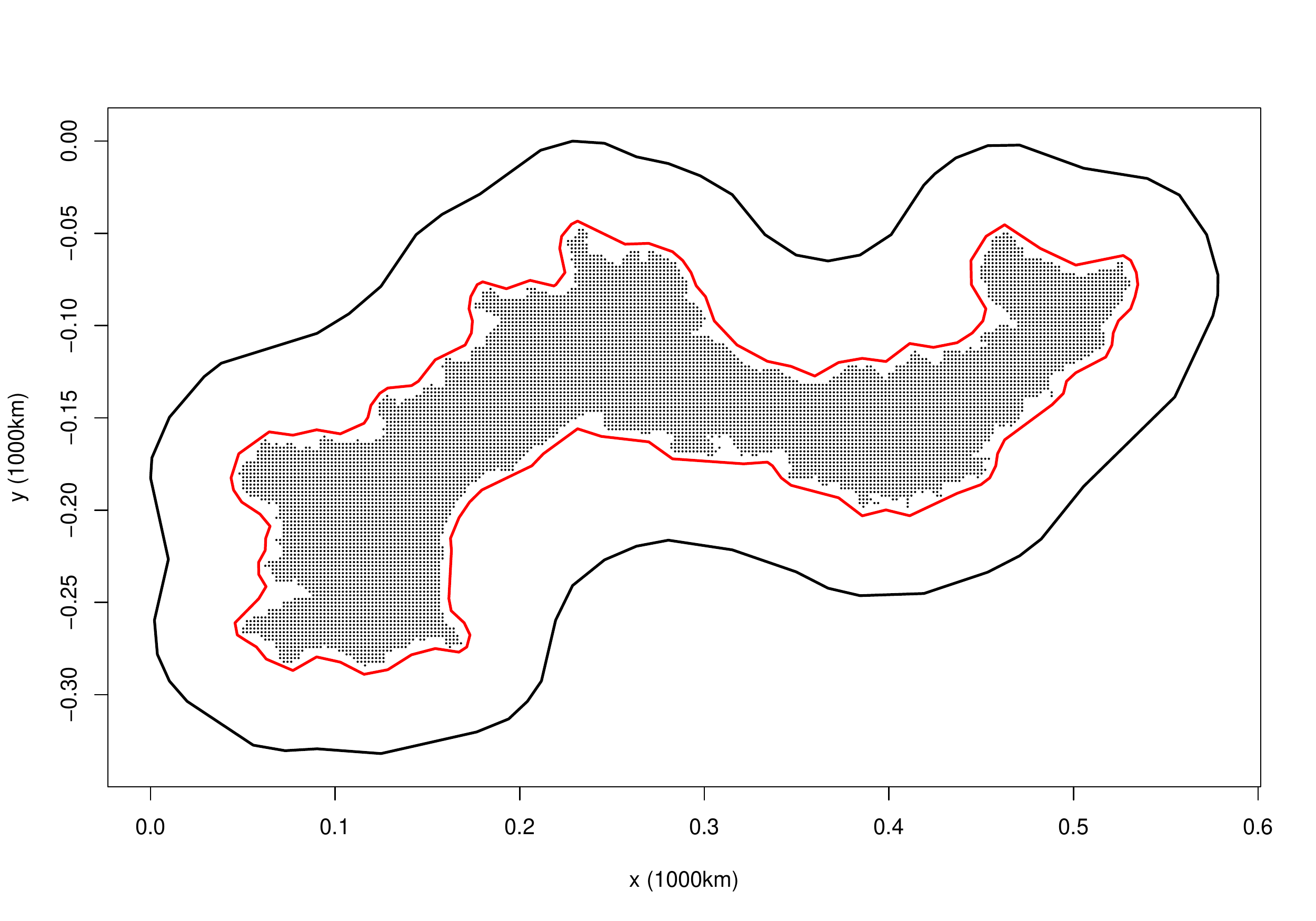}$\quad$
    \includegraphics[width=0.475\linewidth]{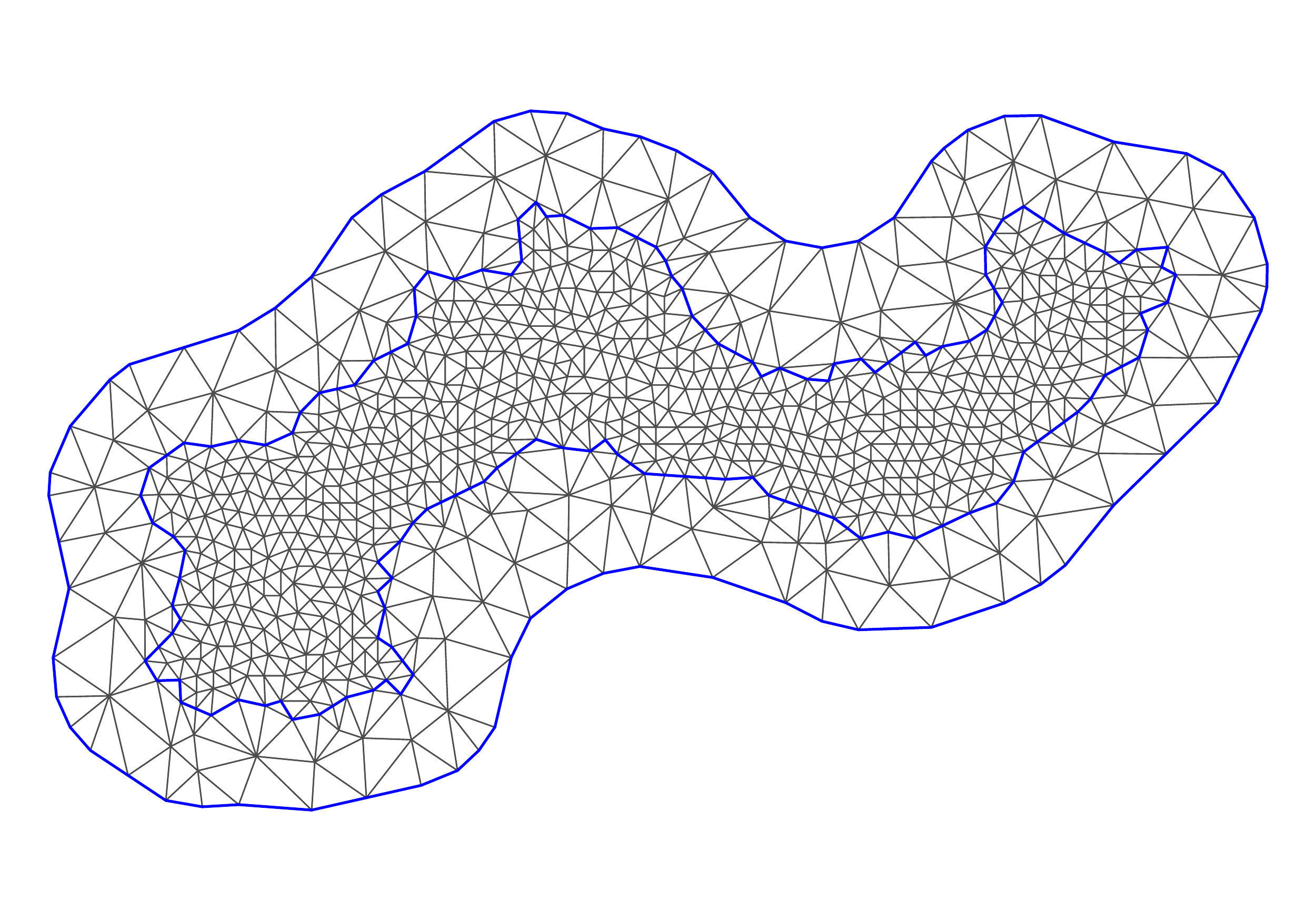}
    \caption{Triangulation mesh. Left: DFCI grid cells and internal and external boundary for the mesh construction (distance units are in $1,000$km). Right: constructed mesh with $816$ nodes. }
    \label{fig:mesh}
\end{figure}

For each of the spatial fields $W(s,a(t))$ for years $a(t)$, we use the well-known  SPDE  approximation of the Mat\'ern covariance function based on a triangulation of space and with regularity parameter fixed to $1$ \cite[see][for theory and practice of the SPDE approach]{Lindgren.al.2011,Lindgren.al.2015,Krainski.al.2018}. The grid of DFCI cells and the triangulation mesh constructed for the spatial effect are presented in Figure~\ref{fig:mesh}.
For spatio-temporal structure in $W(s,a(t))$, we consider four choices: 
\begin{enumerate}
    \item no effect (i.e., $W(s,t)\equiv 0$), or
    \item perfect temporal dependence (i.e., a single spatial random effect $W(s,a(t))\equiv W(s,1)$), or 
    \item independent spatial fields, or
    \item time-stationary autoregression defined as follows:
\begin{equation}\label{eq:star}
   W(s,a(t))=\rho W(s,a(t)-1)+\sqrt{1-\rho^2}\varepsilon_{a(t)}(s), \quad \rho\in (-1,1),
\end{equation}
where $\varepsilon_{a(t)}(s)$ are the spatial Mat\'ern-SPDE innovation fields. 
\end{enumerate}
Temporal structure in models 1 and 2 arises only through the time-varying fixed covariate effects and the seasonal effect, and additionally through the time-replicated spatial effect in model 3.
By adding temporal autocorrelation in model 4, we can capture  the persistence of spatial effects through time, such as subregions that tend to be systematically stronger affected by forest fires than others. Moreover, the temporal dependence could help to appropriately correct relatively smooth nonlinearities with respect to the linear time trend captured through the coefficient $\beta^{\mathrm{time}}$ (if any of such nonlinearities exist). We will estimate the following four hyperparameters, whose posterior distributions will have a strong impact on the structure of the fitted model: the Mat\'ern range parameter (equal to the approximate spatial distance at which correlation $0.1$ is reached); the Mat\'ern variance; the autoregression coefficient $\rho$ (if part of the model); a smoothness parameter for the seasonal effect $f(\mathrm{month}(t))$. Other hyperparameters are the variance parameters of the centered normal prior distributions of fixed effects $\beta_j^{\text{type}}$.  

\section{Bayesian inference using INLA}
\label{sec:bayes}
\subsection{Formulation as a Bayesian regression problem for INLA}
The spatial-temporal resolution of our model corresponds to  DFCI grid cell centers $s_i$, $i=1,\ldots,9562$ and the months $t_j$ of the $24$-year study period (with only $6$ observed months in 2018), $j=23\times 12+6=282$. Consequently, we can consider the counts of fire occurrences $N_{ij}\in\{0,1,2,\ldots\}$ for each space-time cell in the Cartesian grid of $(s_i,t_j)$, $i=1,\ldots,9562,\, j=1,\ldots,282$, and  reformulate the model as
\begin{equation}\label{eq:regmod}
N_{ij}\mid \Lambda(s_i,t_j)\sim \mathrm{Poisson}(4\Lambda(s_i,t_j)), \quad i=1,\ldots,9562,\, j=1,\ldots,282,
\end{equation}
where the constant $4$ means that we use a spatial unit of $km^2$ for $\Lambda$ (note that one DFCI cell covers $4km^2$).
Therefore, the log-linear structure of $\Lambda(s,t)$ with random effects in \eqref{eq:Lambda} necessitates the estimation of a generalized additive mixed model. In our dataset, more than $99\%$ of observed counts $N_{ij}$ are $0$, and the overall number of observed wildfires is $23,309$. If wildfires occur, the occurrence is single (i.e., $N_{ij}=1$) in more than $91\%$ of cases, and less than $0.25\%$ of positive wildfire counts are larger than $4$; only $2$ counts are larger than $9$, and the maximum observed count is $12$. 

The Integrated Nested Laplace Approximation \citep{Rue.al.2009} and its implementation \texttt{R-INLA} in the \texttt{R} statistical software \citep[see][for instance]{Rue.al.2017,Opitz.2017} is a fast and accurate inference tool for high-dimensional generalized additive models where conditional independence of the response distribution arises with respect to  latent Gaussian processes, as it is the case in our Poisson regression model with log-link function in \eqref{eq:regmod}. In the calculation of posterior estimations, it uses an astute  combination of analytical Laplace approximations to perform numerical integration with respect to the latent Gaussian variables, and of numerical discretization to approximate the integrals with respect to  the hyperparameters. Specifically, using INLA to fit log-Gaussian Cox process models with complex nonlinear effects of space, time and covariates in the intensity function  $\Lambda$ has become common practice in recent years \citep[][for instance]{Illian.al.2012,serra2014,GomezRubio.al.2015,Lombardo.al.2018,Lombardo.al.2019b,Lombardo.al.2019}.

\subsection{Choice of prior distributions}
For fixed effect coefficients $\beta_j^{\mathrm{type}}$, we use independent Gaussian prior distributions centered at $0$ with fixed precision $0.1$ (i.e., variance $10$). For the seasonal effect $f(\mathrm{month}(t))$ defined over $1,2,\ldots,12$, we fix a periodic first-order random walk prior such that months $12$ and $1$ are linked. 

We fix moderately informative prior distributions for hyperparameters, which helps stabilizing the estimation procedure by penalizing excessively complex models with unstructured random effects  and by facilitating  convergence of  Laplace approximations. In particular, we make systematic use of penalized complexity (PC) priors \citep{Simpson.al.2017}, which penalize the distance of model components with respect to a relatively simple baseline specification. We refer to  \cite{Fuglstad.al.2018} for the derivation of PC priors for the Mat\'ern covariance where the baseline is a deterministic field with value $0$ everywhere. Effectively, we fix prior distributions such that the prior probability of observing a covariance range below $50km$ and of observing a variance larger than $1$ is $50\%$ in each case. 
Next, we refer the reader to \cite{Sorbye.al.2014} for the definition of prior distributions for the marginal variance of the seasonal first-order random walk $f(\mathrm{month}(t))$; here we fix priors that set the probability of having marginal variance larger than $0.25$ to $50\%$. Finally, for the temporal autoregression coefficient $\rho$ in \eqref{eq:star}, we consider the value $0$ as baseline model, such that a priori no strong temporal persistence of spatial wildfire occurrence patterns is assumed, and the prior distribution puts $50\%$ of its mass to absolute values $|\rho|>0.5$.  

\subsection{Subsampling of intra-year effects}
When estimating the Bayesian regression model \eqref{eq:regmod} with INLA, the response vector of counts has approximately $2.7$ million entries, which has turned out to lead to excessively high memory requirements and to numerical instabilities when running INLA, even on machines with $128Gb$ of memory.
Section $8.4$ in \cite{Krainski.al.2018} presents strategies consisting in both aggregating the events as well as lowering the spatial or temporal mesh resolution to decrease the computational time for large point process datasets. In our case, this methodology would come along with a deterioration of results for structures at relatively small scales, and it is also problematic with respect to covariates, which have to be aggregated to larger spatial units than DFCI pixels. By consequence, we have instead opted for an approach of intra-year aggregation of pixels with count $0$. It causes only moderate loss of information as compared to the original data. We have devised our subsampling procedure as follows.
It exploits Poisson additivity, i.e. the fact that $N_1+N_2\sim \mathrm{Poisson}(\lambda_1+\lambda_2)$ if the two count variables $N_1\sim \mathrm{Poisson}(\lambda_1)$ and $N_2\sim\mathrm{Poisson}(\lambda_2)$ are independent. We use this mechanism to replace the $0$ counts observed for the same configuration of DFCI cell and year by a single $0$ observation, i.e. by adding up all the zeros, and we fix the covariates of this new observation by drawing it at random from the original observations.
Therefore, positive counts  $N_{ij}>0$ and their covariate configurations are kept in the model without modification. Next, for each year and each DFCI grid cell $i_0$, we consider the months without fire occurrences. If there are $1\leq k\leq 12$ of such months, we sample at random one of these months $j_0$ and keep only the value $N_{i_0j_0}=0$ and the covariate configuration of this month, while we discard the $k-1$ other months from the dataset. Moreover, in the regression model \eqref{eq:regmod}, we set $N_{i_0j_0}\mid \Lambda(s_{i_0},t_{j_0})\sim \mathrm{Poisson}(4k\Lambda(s_{i_0},t_{j_0}))$ to account for the $k-1$ removed response values. 
This subsampling scheme reduces the length of the response vector to approximately $250,000$, and has led to a stable estimation procedure with \texttt{R-INLA}. We can here argue that the loss of information from the original data is limited since the sampled grid cells $i_0$ still provide a good coverage of the study region for each month, with very similar weather conditions between close DFCI grid cells. Moreover, the set of covariate configurations associated to observations with no fire occurrence, as determined through the subsampling procedure, remains very large and  can still be considered as representative for the set of original covariate configurations. 

\section{Results}
\label{sec:res}
For the results reported in the following, we focus on the most ``complex" model 4 presented in Section~\ref{sec:StLGCP}, where the yearly spatial random effects are linked through temporal autocorrelation. Estimations of fixed covariate effects and of the seasonal effect with monthly resolution have been very similar in terms of magnitude and significance between the models 1 to 4  with different specifications of the space-time Mat\'ern random effect. For a quantitative assessment of relative improvement in goodness-of-fit of model 4 with respect to the models 1 to 3, we report the difference in  marginal likelihood values between  models 1 to 3  with respect to model 4; these values are $-4.2\times 10^3$ (model 1), $-3.0\times 10^3$ (model 2) and $-1.2\times 10^3$ (model 3). Therefore, we see that each of the three modeling extensions (single spatial random effect;  independently replicated spatial random effect; spatial random effect with temporal autoregression) with respect to the baseline model (fixed effects, seasonal monthly effect) leads to a substantial augmentation of marginal likelihood by more than $1,000$. 

Regarding the results of model 4, we first study the posterior mean estimates of the four hyperparameters, with estimated standard errors indicated in parentheses (Table~\ref{tab:hyperpara}). The spatial range is estimated at $20km(1km)$, such that fires tend to ignite in a spatially clustered way, although at relatively small scales. In the construction of the spatial triangulation for discretizing the Mat\'ern SPDE field, we have set a maximum edge length of $12.5km$; therefore, we can suppose that our model can still differentiate between the situation of no spatial dependence and weak but significant spatial dependence within the estimated range of $20km$. Moreover, the standard deviation of the spatial fields is estimated at $1.36$, such that there is substantial variability in the space-time intensity that cannot be explained by the fixed and seasonal effects. The estimate of the temporal autocorrelation coefficient $\rho$ in \eqref{eq:star} is $0.89(0.009)$, such that spatial clustering patterns seem to persist quite strongly over consecutive years. Finally, the precision (i.e., 1/variance) of the seasonal effect with monthly resolution is estimated at $2.4(1.0)$.
{
\begin{table}[]
    \centering
    \begin{tabular}{c|cc}
\textbf{Hyperparameters} & \textbf{Posterior mean} &\textbf{Standard deviation}\\
\hline
Spatial range ($km$) & 20 & 1.0 \\
Standard deviation of the spatial field & 1.36 & 0.035 \color{black} \\
Temporal autocorrelation & 0.89 & 0.009 \\ 
Precision of the seasonal effect & 2.4 & 1.0\\ 
\hline 
\end{tabular}
\caption{Estimates of the hyperparameters of the spatial Mat{\'e}rn-SPDE innovation field $W$ in model 4.}
    \label{tab:hyperpara}
\end{table}
}

Moreover, we underline a significant negative linear time trend in the log-intensity; see Table~\ref{tab:covars}. Indeed, it corresponds to an almost $40\%$ drop in the point process intensity when comparing the beginning and the end of the $24$-year study period. In the context of a warming climate with a tendency towards higher positive temperature anomalies and more arid climate in the Mediterranean basin, this contrary effect can be interpreted as a consequence of the combination of increased vigilance and a set preventive measures coordinated by the competent local authorities, firefighters and forest wardens. Nevertheless, observing extremely high yearly  fire counts is not excluded for recent years; for instance, we observe very large numbers of wildfires in 2017.

In the following subsection, we present detailed results and interpretations for the  significant fixed effects with respect to the other $33$ covariates.

\subsection{Influence of season, weather and interfaces of forest to human activity}

Recall that our model 4 has monthly temporal resolution over the spatial DFCI grid, with covariates for weather anomalies obtained through spatiotemporal kriging of  temperature and precipitation. 
Land cover covariates at the DFCI grid resolution are included through means and standard deviations of high-resolution land cover data. In particular, including the variability of the coverage inside DFCI cells allows us to identify the effect of the heterogeneity of the land cover distribution on  wildfire outbreak. The standard deviation can also be interpreted as a measure that approximately quantifies the length of the interface between a given land cover type and the other types. Moreover, to study wildland-to-urban interfaces, we have introduced two additional synthetic covariates, one for joint forest and building coverage, the other for joint forest cover and road length.

Table~\ref{tab:covars} presents the model coefficients significantly different from 0 associated to each covariate.
Model coefficients higher (respectively lower) than $0$ indicate that a high covariate value tends to lead to an increase (resp. decrease) of the fire occurrence probability. In the sequel, each number in parentheses refers to the id number of the corresponding covariate in Table~\ref{tab:covars}.

The land cover covariates expressed through the mean over the DFCI cell which contribute to increase the probability of occurrence of a forest fire are the following: temperature anomaly (2), road length (10), proportion of conifers (14), slope (5), and  area of regulated tourist areas (18). On the other hand, high values of the following factors significantly decrease this probability: altitude (3), precipitation anomaly (1), water coverage (20), the building covarage (6), length of secondary roads (12), and  length of paths (8).

For the French Mediterranean area, our results precisely quantify the current knowledge about factors that tend to favor or limit the occurrence of a forest fire. We believe that altitude, with the most significant coefficient, already summarizes part of such global information. Indeed, low altitude levels can be found near the coastline where the climate is the most Mediterranean, hot and dry, with highly flammable plant species (conifers) and a strong human presence (buildings, roads, tourism), whereas at higher altitudes temperatures are lower, precipitation is higher, vegetation is less present and less favorable to ignition, and human activity is lower or more strongly supervised in the case of tourism. The total length of roads can be seen as a proxy for human presence, while conifers are very present along the coast in the Provences-Alpes-C\^ote d'Azur region and represent a highly flammable tree species. The slope is an important factor in the mountains and provides complementary information with respect to  altitude; indeed, on the Mediterranean coast we can find areas with low elevation but with steep slopes (creeks, valleys) and a lot of tourism.  Slope is known to be a factor of propagation of forest fires, therefore it is logical to find this covariate significantly positive. The significance of the regulated tourist zone covariate might be explained by the fact that, despite the efforts of conservation and prevention of the forested areas, touristic pressure is so high in this region of the South of France that it increases the risk of wildfire occurrences. DFCI areas with a very high proportion of buildings (urban areas) or water naturally have a much lower level of forest fire exposure. According to our model, the presence of many secondary roads and paths tends to limit the occurrence of a forest fire. This characteristic contrasts with an opposite effect observed for the total length of all roads, highlighting the major impact of primary roads.

Next, we analyze the interface effects of forest-to-building and forest-to-paths, both of which are significant. The forest-building covariate (21) increases the fire occurrence intensity, while the forest-to-paths factor (22) leads to a decrease. Forest-to-building interfaces concentrate the main cause of wildfire outbreaks: human activity in a forest environment. The risk reduction owing to the forest-to-paths factor is more difficult to explain, but may be due to the fact that the presence of small dirt roads and paths does not necessarily coincide with fire-hazard-prone human activities. 

We now study the coefficients of the standard deviation covariates that are significantly different from $0$. A first group of variables  increases the probability of fire occurrence: secondary road length (13), forest cover (16), path length (9), building cover (7), shrubland (19), conifers (15) and moorland (17);  while a second group of variables are associated to a decrease of wildfire intensity: road length (11), altitude (4). Overall, these effects are in line with our general understanding that locally heterogeneous environments, where human activity often coincides with presence of combustible material, favor the outbreak of wildfires. The effects of the different road types are not always easy to disentangle, but our model shows that allowing for the interplay of average road lengths and variances for different types considerably improves the goodness-of-fit. 

\begin{figure}
    \centering
    \includegraphics[width=.6\linewidth]{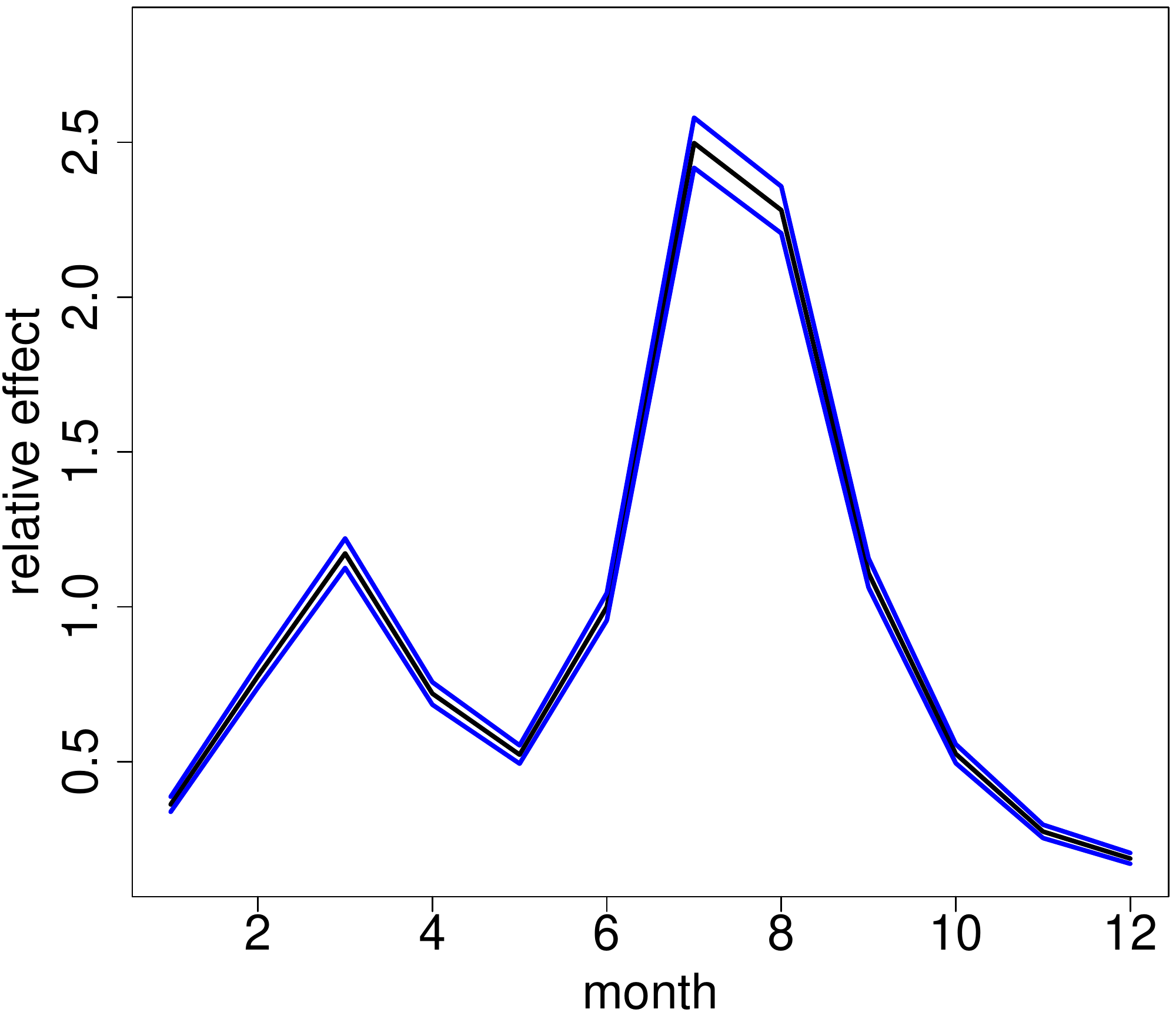}
    \caption{Estimated seasonal effect. The plot shows the posterior mean (black curve) of the odds ratio with the month of June as reference (i.e., with its value scaled to $1$). Blue curves show the symmetric $95\%$ credible interval.}
    \label{fig:season}
\end{figure}

Finally, we show the residual seasonal effect in  Figure~\ref{fig:season}. An explication of its two-peak structure comes from the vegetation cycle and the flammability of vegetation during different seasons of the year. At the end of winter and the beginning of spring (months 2 to 4 approximately), vegetation is still relatively dry and new plant shoots and leaves only start to appear, such that fires break out relatively easily. Fire outbreak risk then drops during the following spring months with relatively high precipitation and relatively ``green" vegetation. Finally, the highest peak arrives during the summer months (July and August) where extremely dry conditions, dry vegetation and very high touristic activity coincide. 

Overall, our Bayesian modeling approach allows for a global and simultaneous consideration of all these factors in space and time, it leads to a prioritization of such factors,  and it provides a methodology that can easily be reproduced or updated for other covariates,  other regions or more recent environmental and fire occurrence data.

\begin{table}[]
    \centering
    \begin{tabular}{l|l|l|l|rr}
\textbf{Id} & \textbf{Type} & \textbf{Category} & \textbf{Covariate} & \textbf{Estimate} &\textbf{CI}\\
\hline
1 & Climate & & Precipitation (square root) & -3.15 & [-3.66,-2.65] \\
2 &         & & Temperature anomaly & 0.09 & [0.08,0.1]\\
\hline
3 & Land & Topography & Altitude (av.) & -1.48 & [-1.64,-1.33]\\
4 &      &            & Altitude (sd) & -1.56 & [-2.66,-0.46]\\
5 &      &            & Slope (av.) & 1.2 & [0.5,1.9]\\
     \cline{2-6}  
6 &      & Urban & Building cover (av.) & -5.21 & [-6.71,-3.7]\\ 
7 &      &       & Building cover (sd) & 2.71 & [1.38,4.04]\\ 
8 &      &       & Path length (av.) & -0.89 & [-1.64,-0.14]\\ 
9 &      &       & Path length (sd) & 1.49 & [0.83,2.15]\\ 
10 &     &       & Road length (av.) & 2.45 & [2,2.91]\\ 
11 &     &       & Road length (sd) & -1.87 & [-2.45,-1.29]\\ 
12 &     &       & Secondary road length (av.) & -1.28 & [-1.81,-0.76]\\ 
13 &     &       & Secondary road length (sd) & 2.69 & [2.11,3.27]\\ 

     \cline{2-6}  
14 &     & Vegetation & Coniferous cover (av.) & 0.36 & [0.17,0.55]\\ 
15 &     &            & Coniferous cover (sd) & 0.29 & [0.04,0.54]\\ 
16 &     &            & Forest cover (sd) & 0.77 & [0.49,1.04]\\ 
17 &     &            & Moorland (sd) & 0.21 & [0,0.43]\\ 
18 &     &            & Protected zone cover (av.) & 0.14 & [0.05,0.22]\\ 
19 &     &            & Shrubland (sd) & 0.33 & [0.05,0.6]\\ 
20 &     &            & Water (av. coverage) & -1 & [-1.21,-0.8]\\ 
     \cline{2-6}  
21 &     & Interfaces & Forest cover+building cover & 4.53 & [2.27,6.79]\\
22 &     &            & Forest cover+paths & -2.54 & [-4.06,-1.02]\\ 
\hline
23 & Time & & Time & -0.48 & [-0.91,-0.05]\\ 
\hline 
\end{tabular}
\caption{Significant fixed covariate effects at the $95\%$ level (based on the credible intervals indicated in brackets in the column CI) for the model 4 in Section~\ref{sec:StLGCP}. The effects are presented (with an id number) by type and category, in alphabetical order, with their posterior mean estimate and CI. Non-significant fixed covariate effects are not reported. In the covariate names, ``average" indicates the use of average values over covariate pixels for each DFCI cell, while ``standard deviation" refers to the use of the standard devation of the covariate values; see Section~\ref{sec:data} for details.}
    \label{tab:covars}
\end{table}

\subsection{Intensity mapping}
\begin{figure}
    \centering
    \includegraphics[width=.375\linewidth]{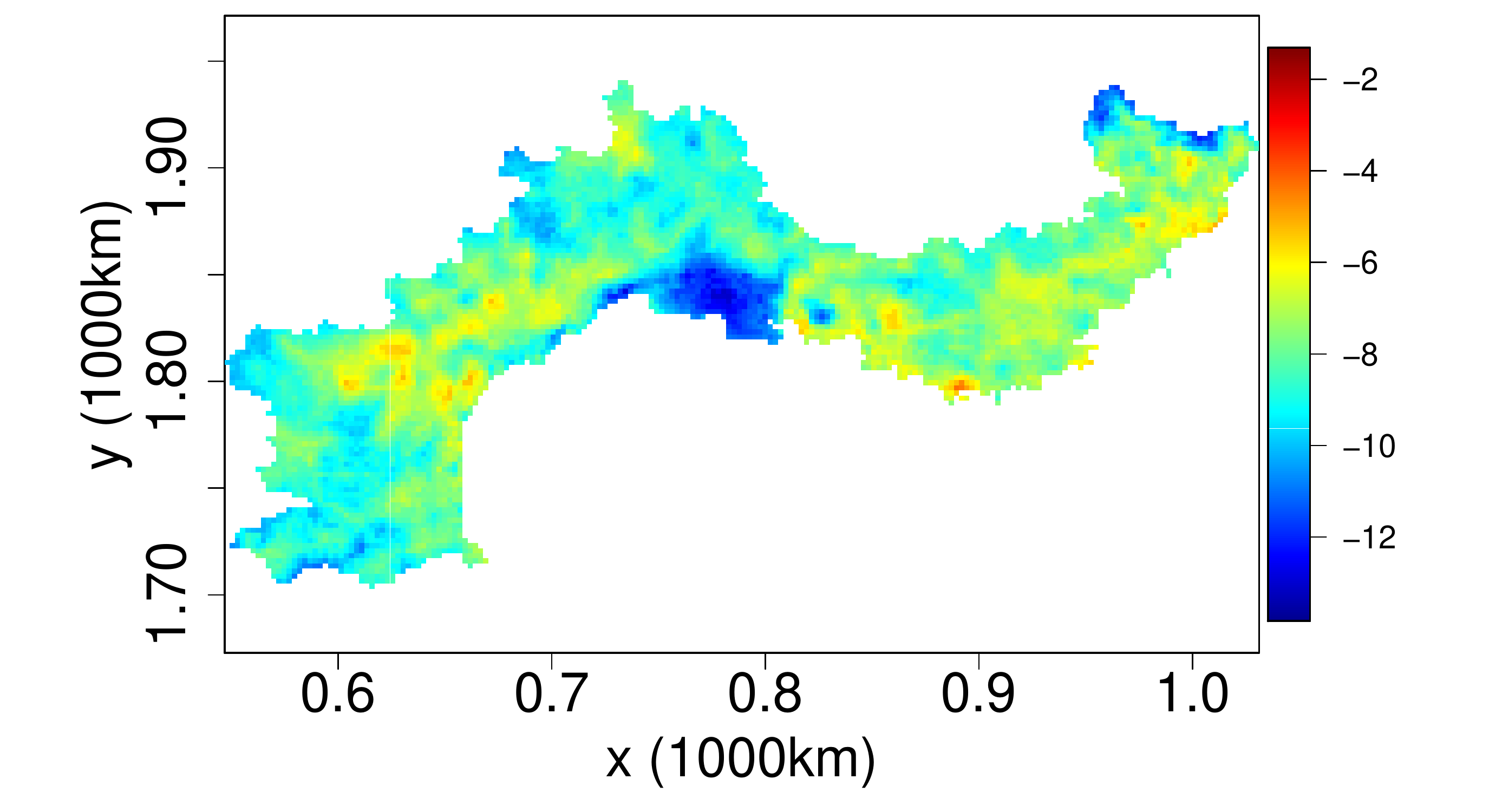}
     \includegraphics[width=.375\linewidth]{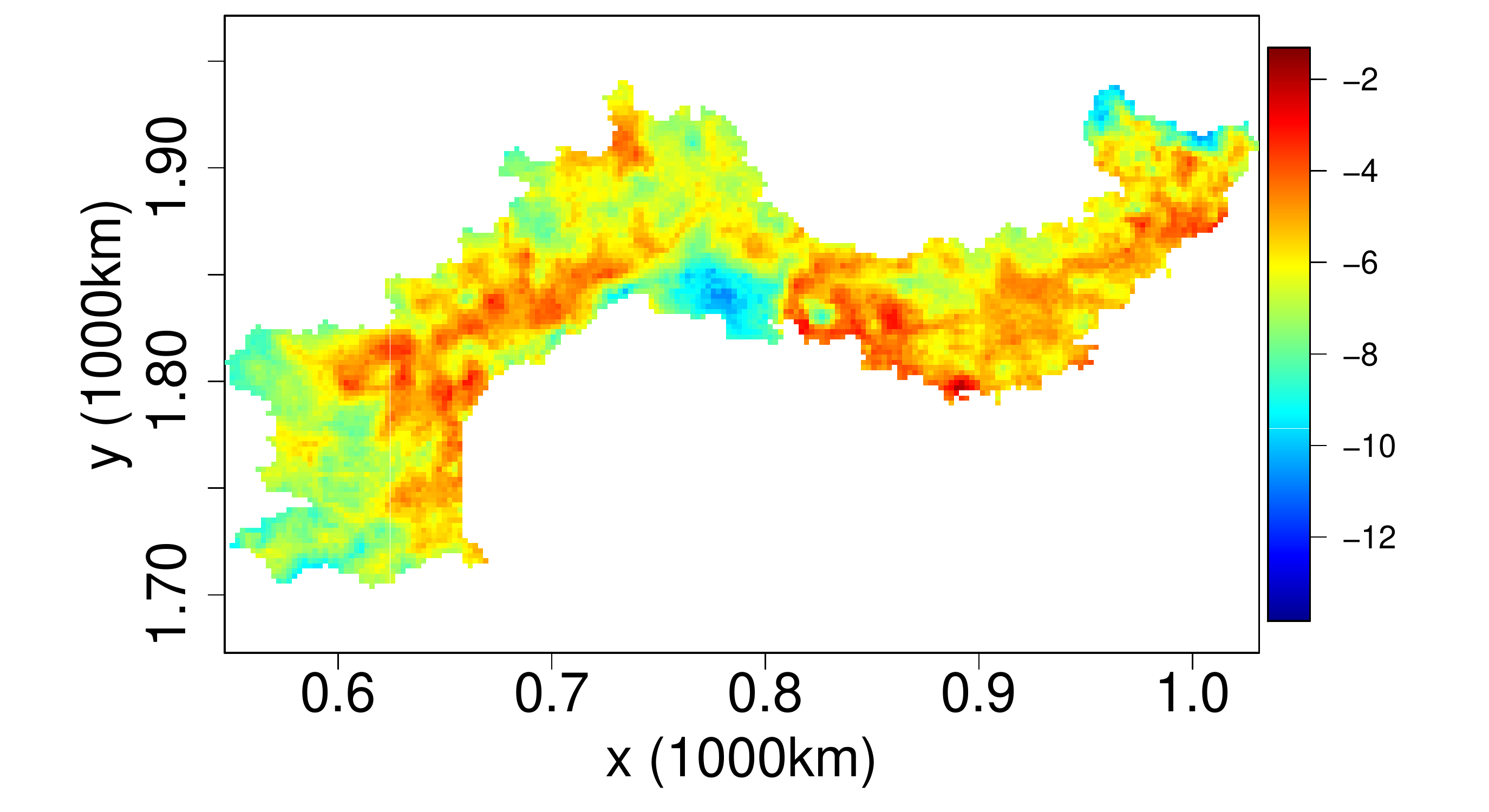}
     \includegraphics[width=.375\linewidth]{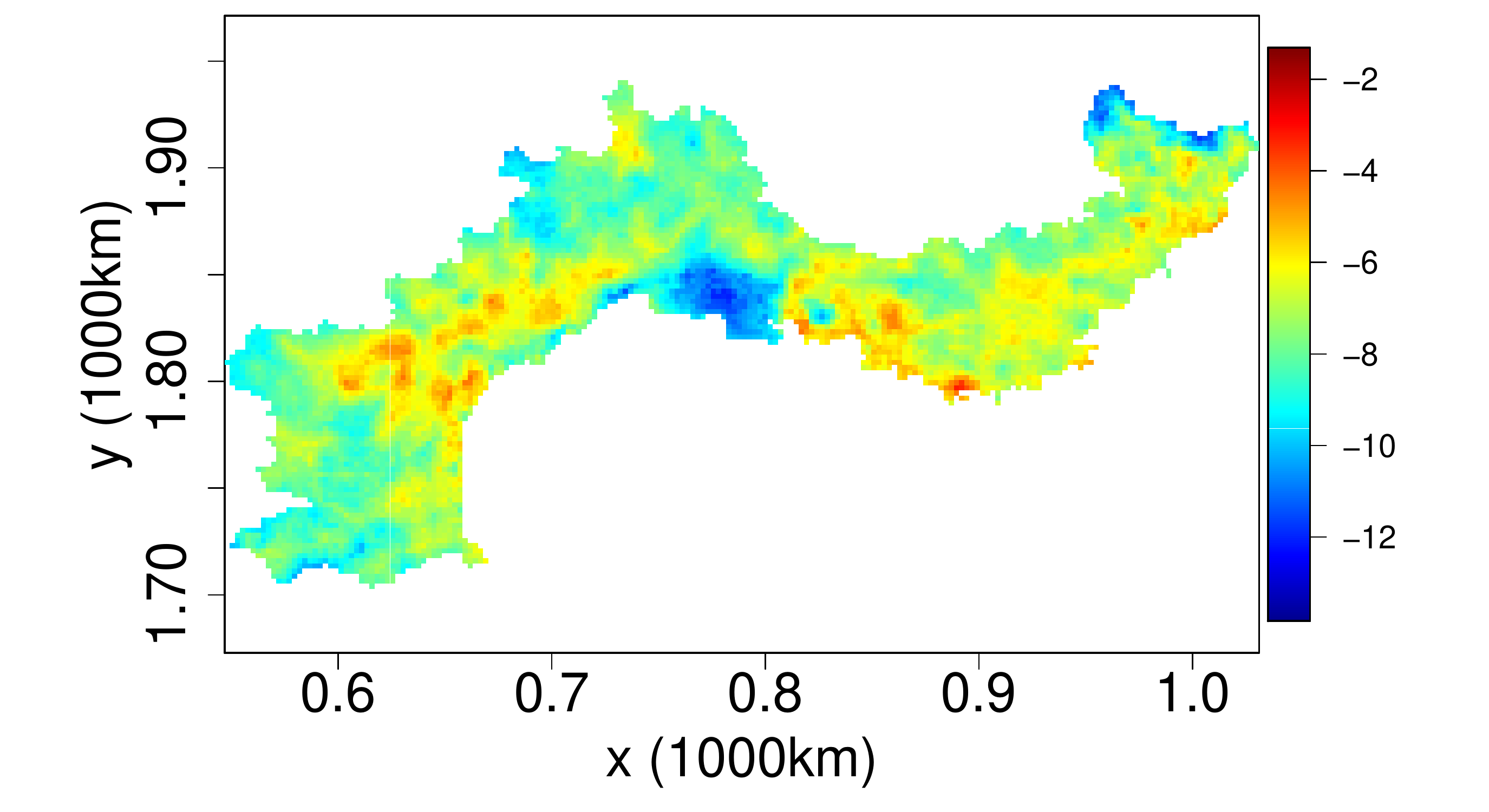} 
     \includegraphics[width=.375\linewidth]{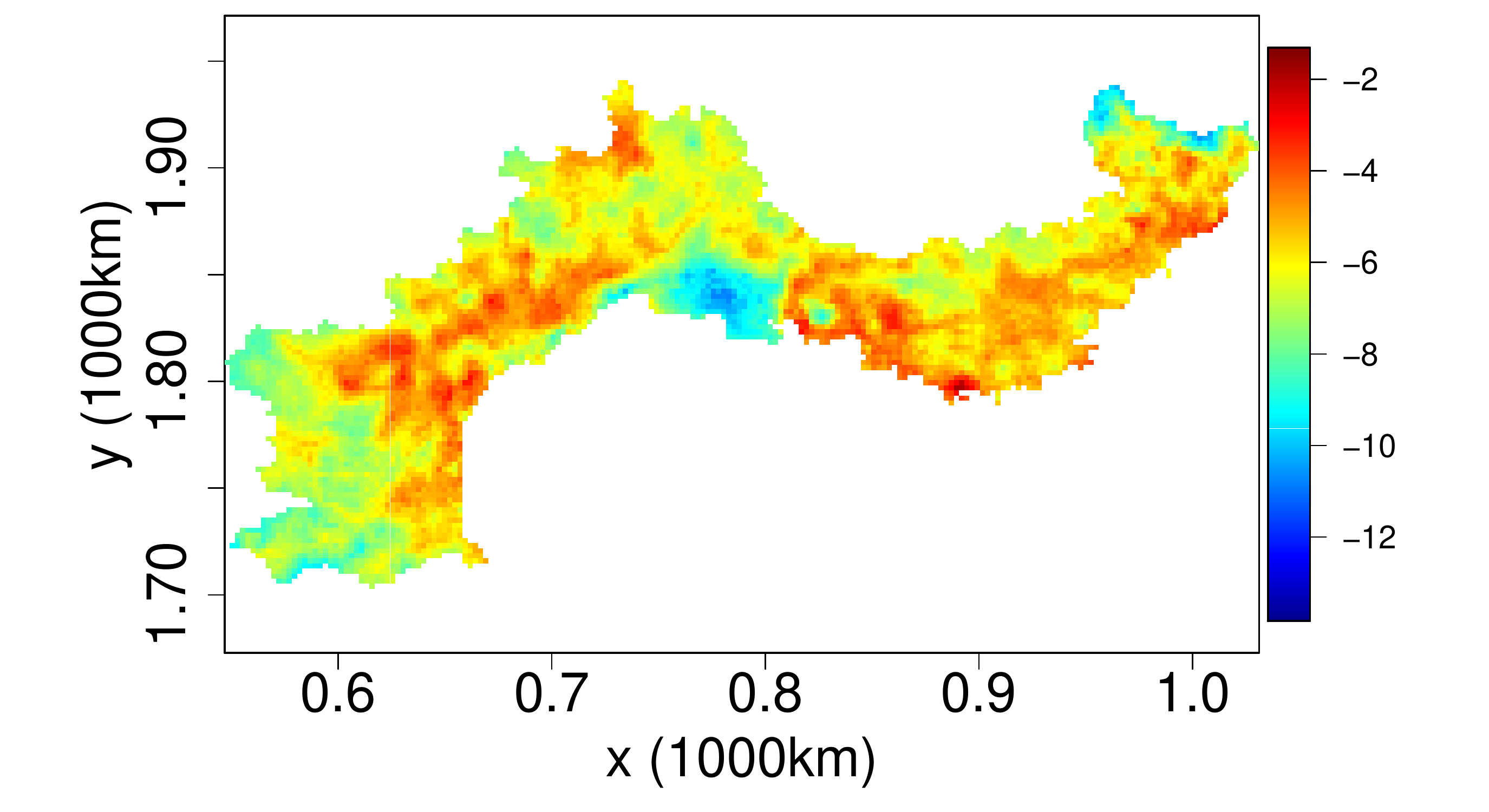} \includegraphics[width=.4\linewidth]{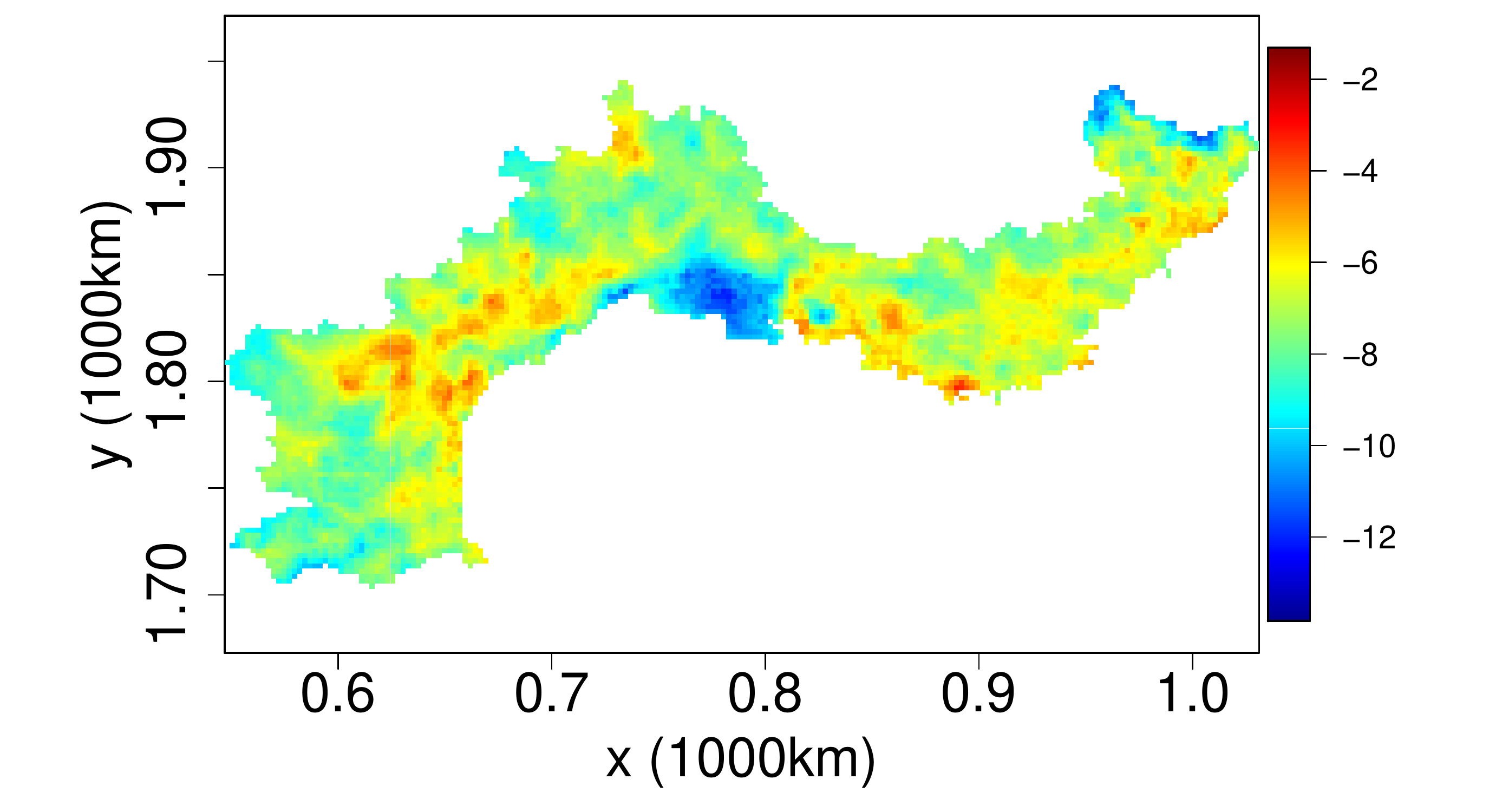}
     \includegraphics[width=.375\linewidth]{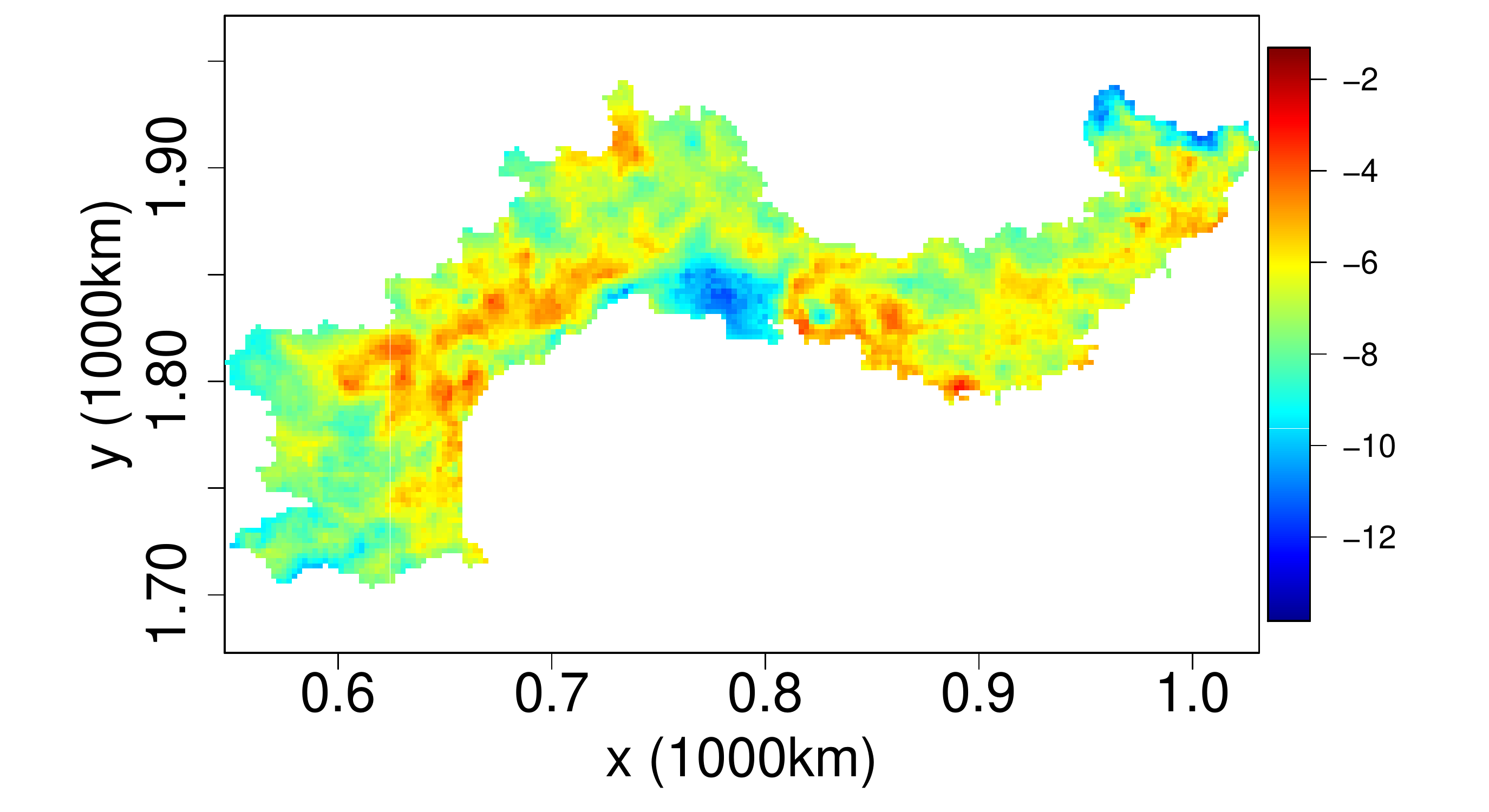} 
      \includegraphics[width=.375\linewidth]{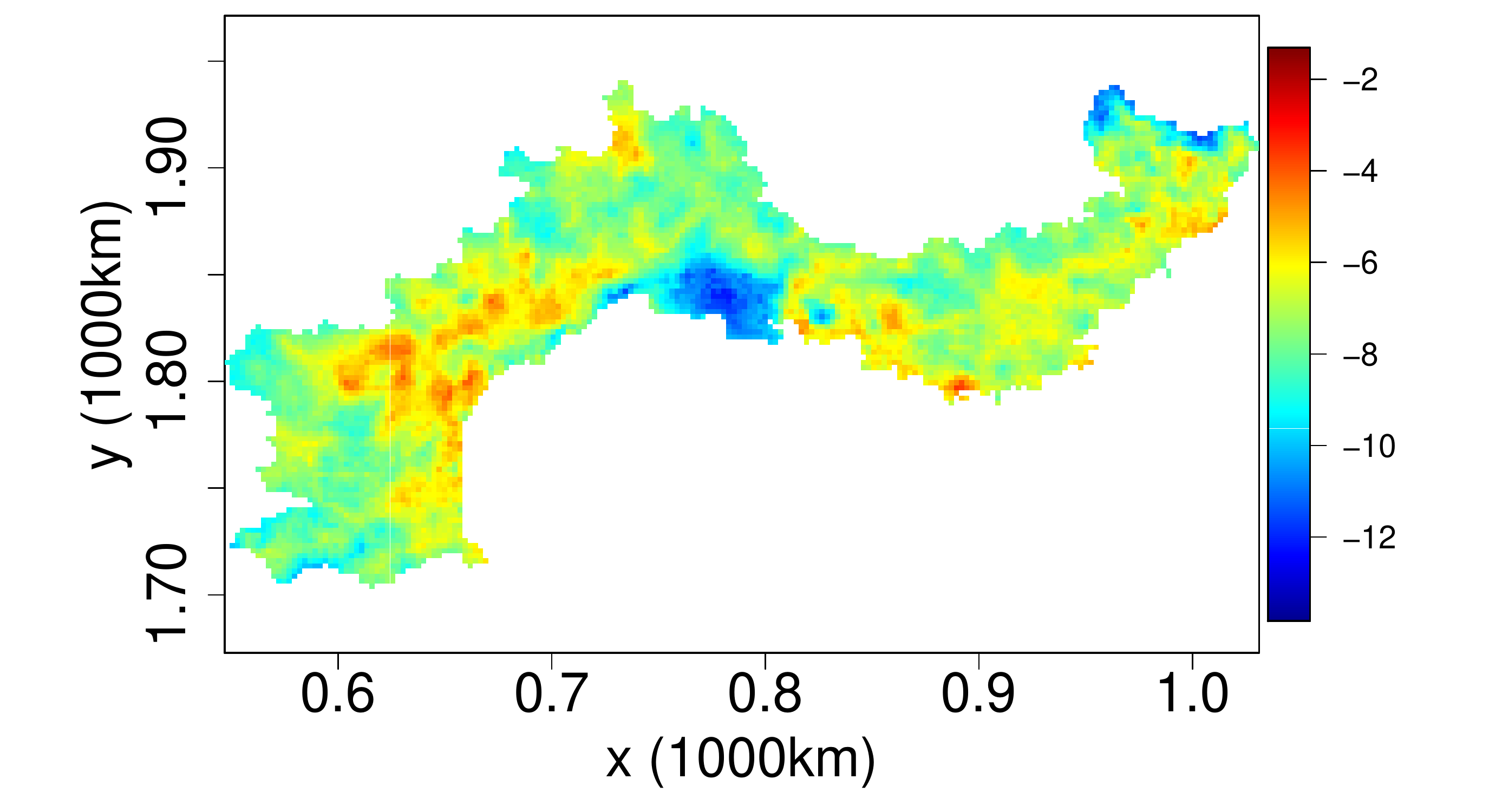}
     \includegraphics[width=.375\linewidth]{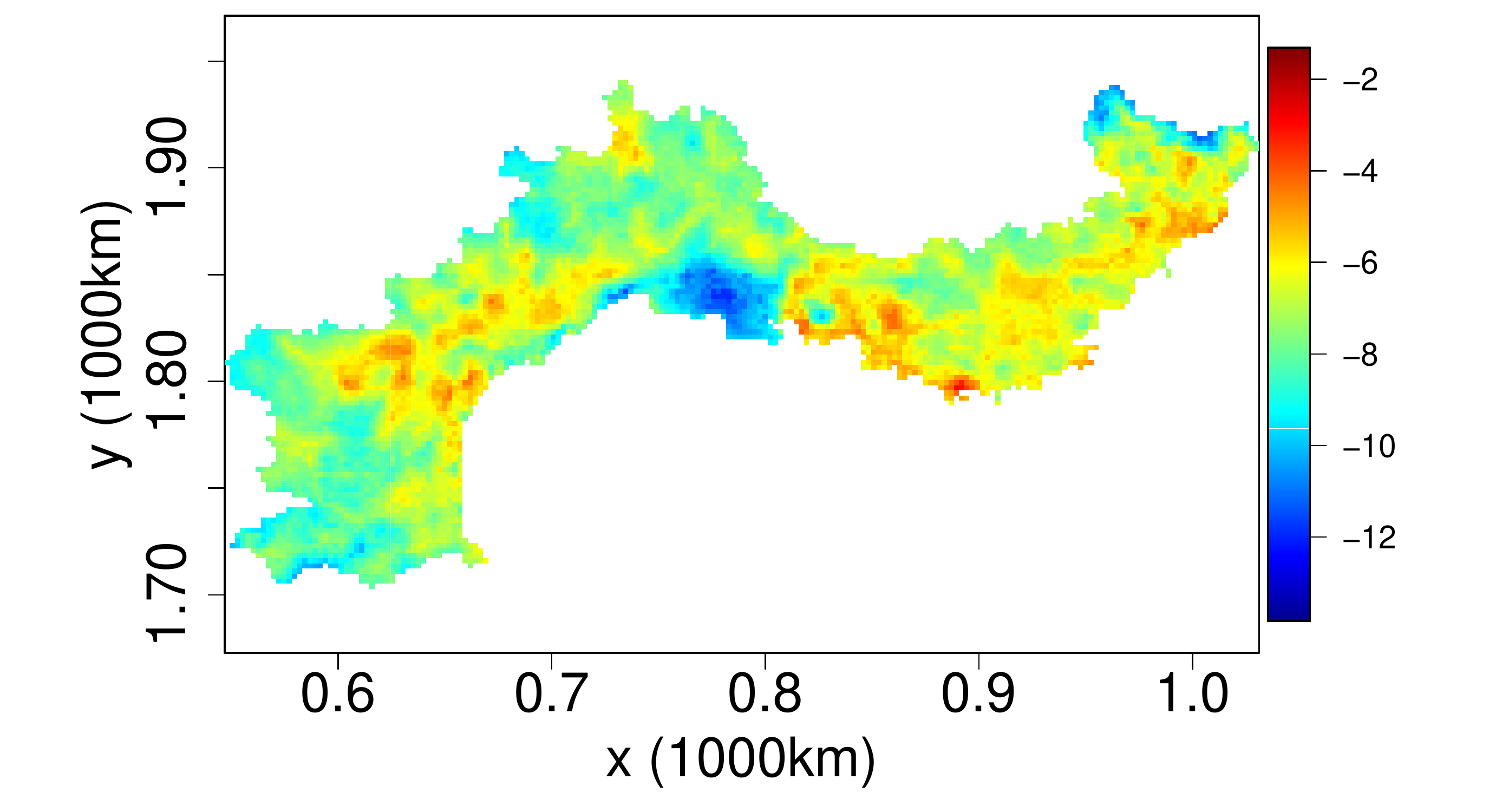}
  \includegraphics[width=.375\linewidth]{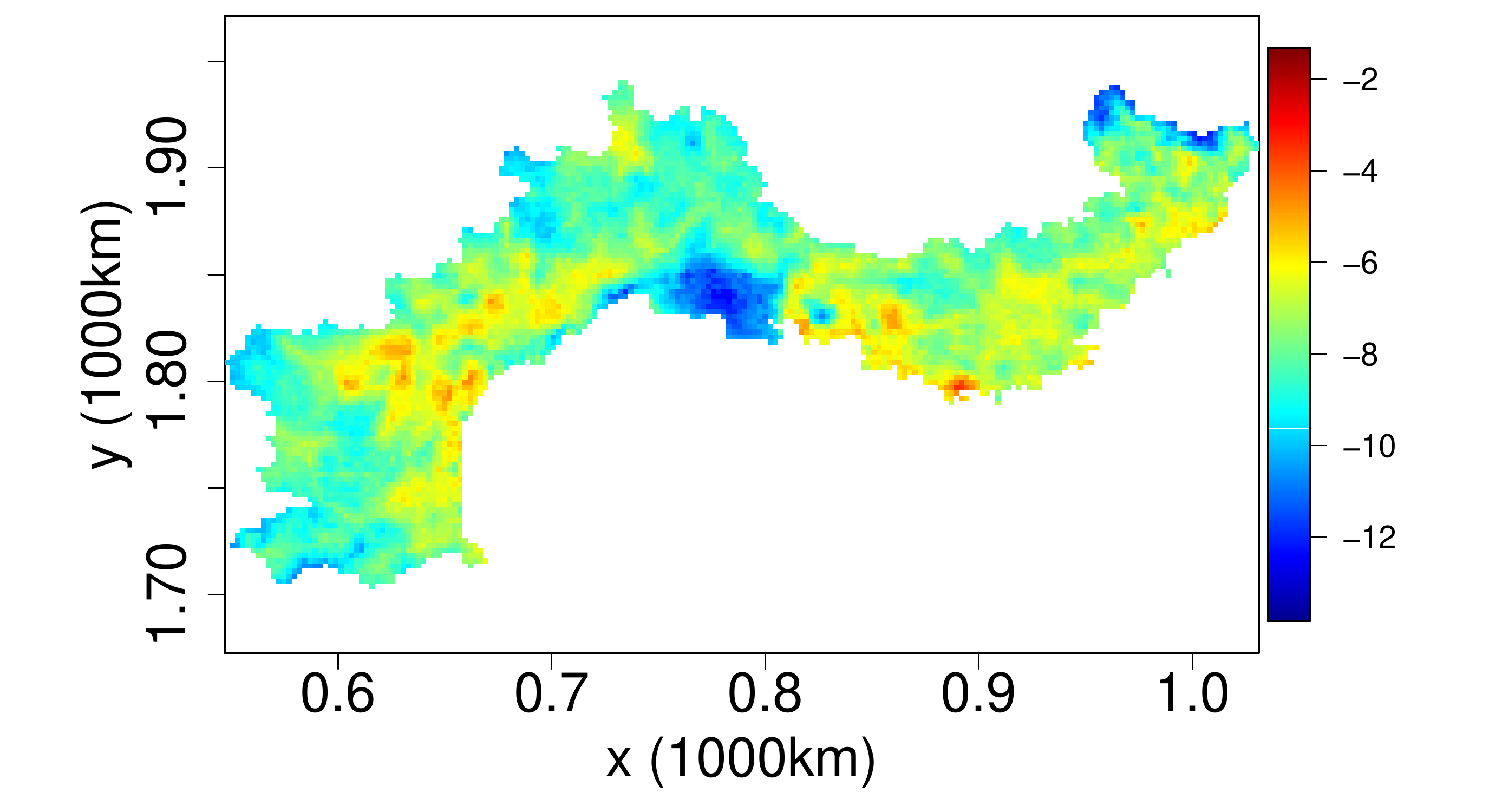}
\includegraphics[width=.375\linewidth]{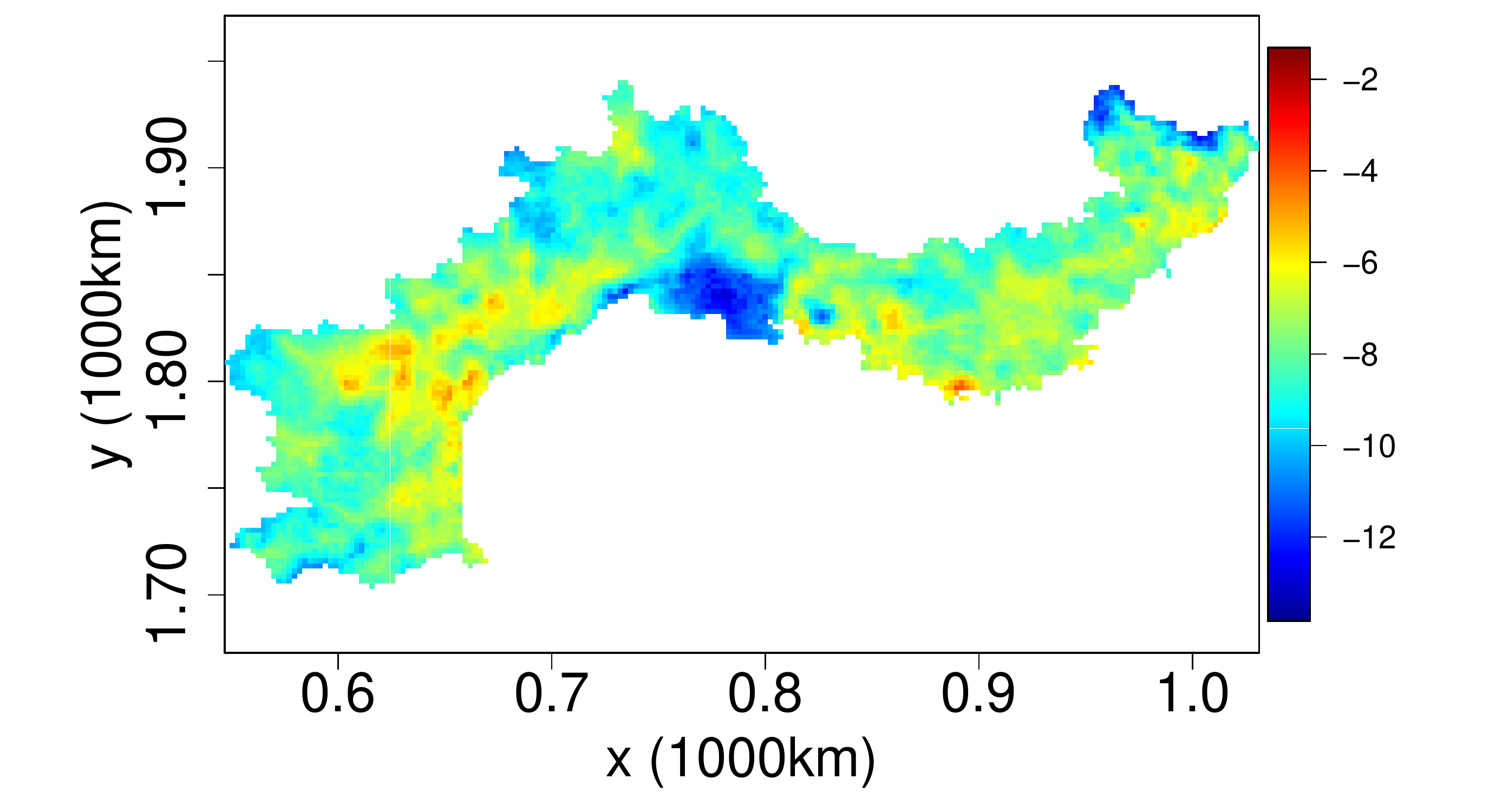}
  \includegraphics[width=.375\linewidth]{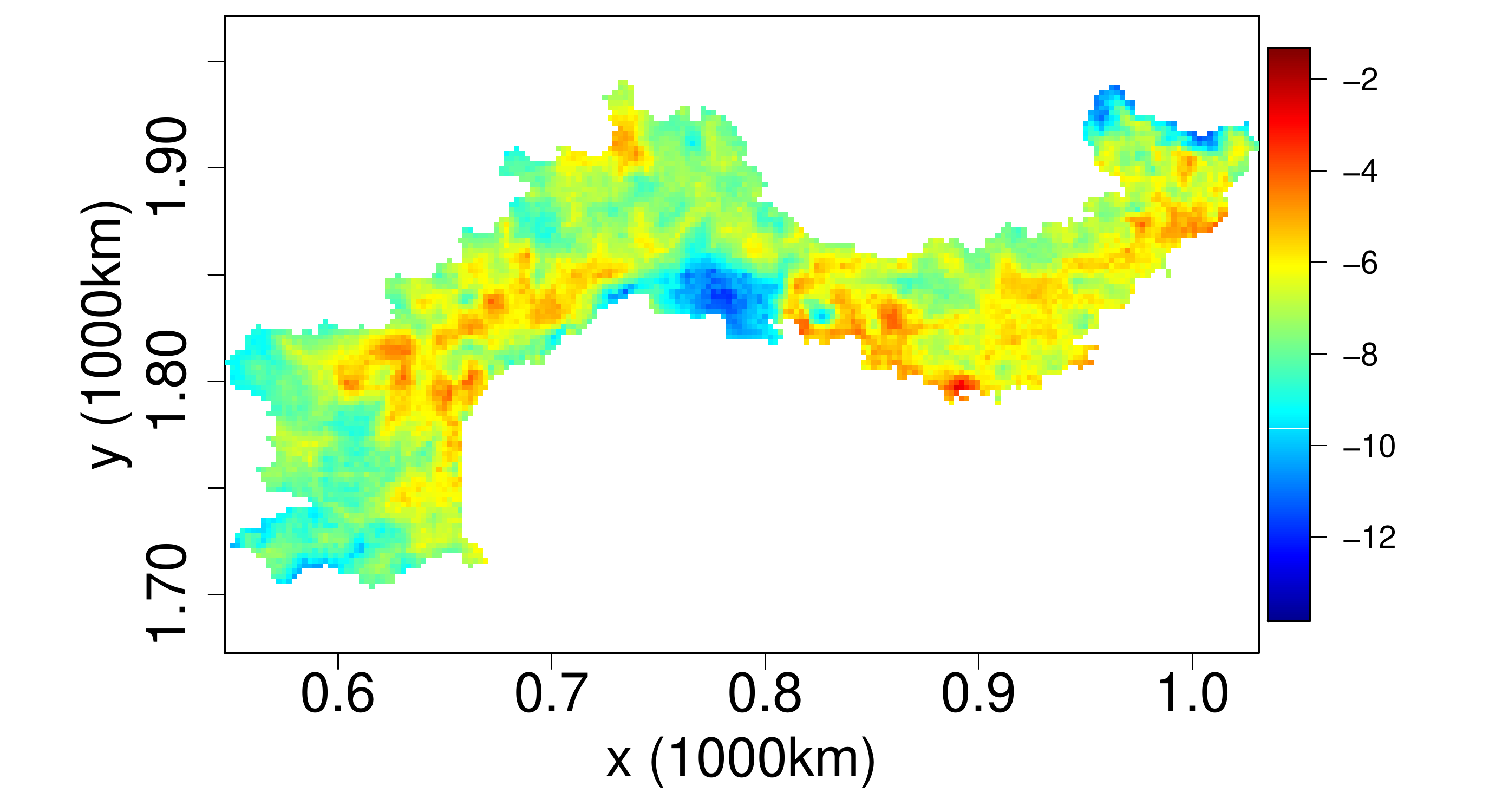}
\includegraphics[width=.375\linewidth]{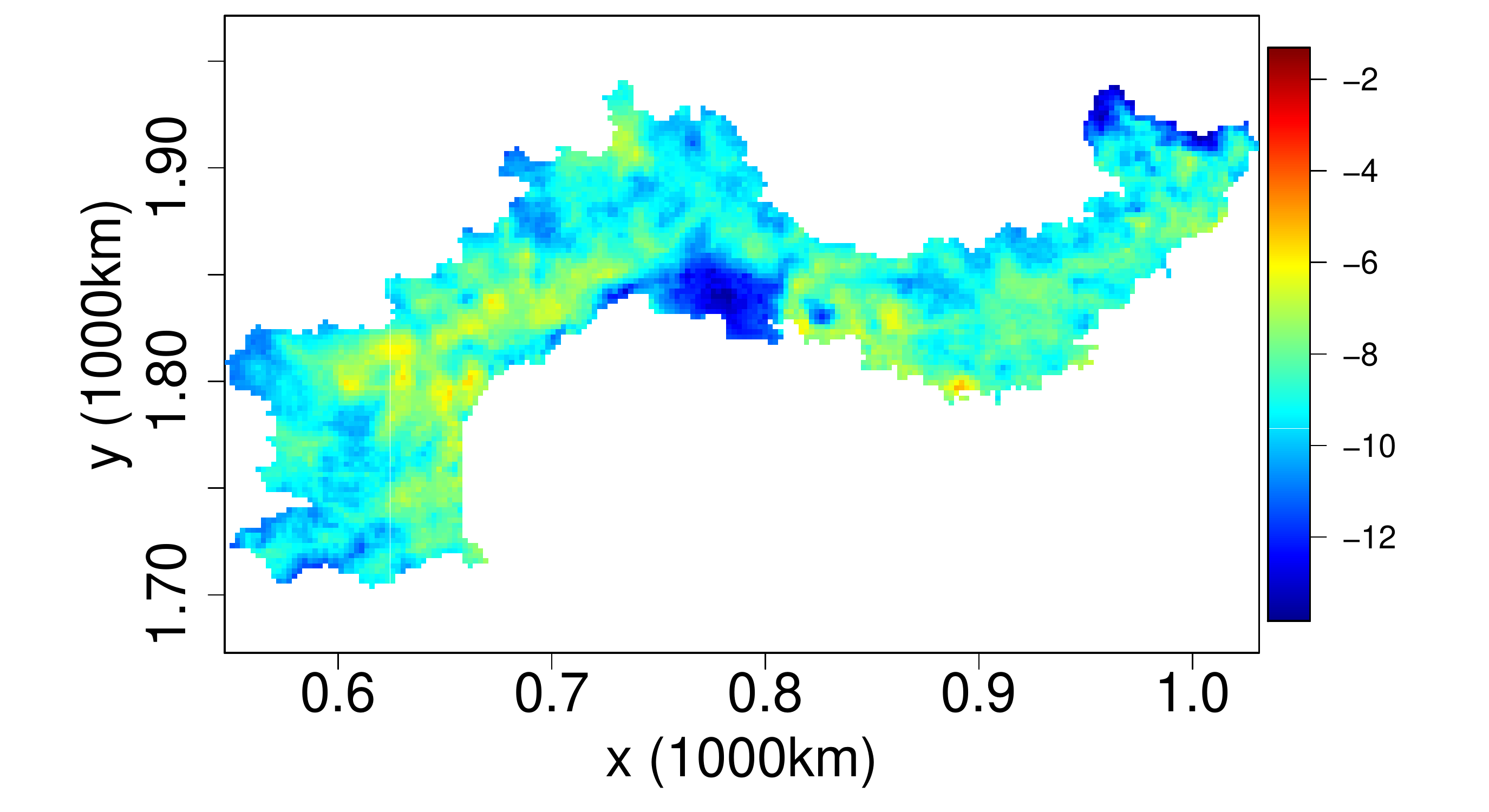}
    \caption{Estimated log-intensity functions $\log(\Lambda(s,t))$ for the months of January to December of 2017, with spatial unit $1/(km^2)$. First column: months January to June from top to bottom. Second column: months July to December from top to bottom. }
    \label{fig:map2017}
\end{figure}

In Figure~\ref{fig:map2017}, we illustrate the posterior mean of the latent log-intensity $\log(\Lambda(s,t))$ estimated from our model 4 by showing maps for the $12$ months of the year $2017$, which has shown the highest occurrence numbers ($1,263$) in the period after 2003. Clearly, the combination of weather influence and  the ``residual" seasonal effect leads to substantial differences between the 12 maps, with rather high values for summer months (July, August, September) and rather low values for winter months (December, January, February).

\section{Discussion and conclusion}
\label{sec:conc}
In the context of a changing climate, wildfires will remain a major challenge for the human societies and natural ecosystems around the Mediterranean. 
The significant covariate effects revealed by our model for mainland Southern France point out the important contribution of unusually high temperatures and low precipitation amounts to increased fire occurrence risk. Moreover, we observe a strong effect of land use, especially of areas where human activity (agriculture, recreation, tourism) takes place in the presence of forest cover. In particular, DFCI cells with high average forest cover alone were not found to be exposed more strongly to fire risk in a significant way; rather, the presence of buildings and forest together, the dominance of coniferous trees, and a fragmented forest cover have been identified  as important factors (among others) contributing to increased fire occurrences. 
Our model can be seen as a first important step towards short-term operational forecasting of fire occurrence intensity, for instance at a weekly scale.  Indeed, if data reaching until the month of December of the preceding year $n$ have been used for fitting our model, and if weather scenarios are available for the current year, it is possible to provide an intensity forecast for the current year. With \texttt{R-INLA}, this solution is simple to implement by specifying the covariates (without missing data) and the count values (missing) for the current year. 
Statistical inference with even higher-dimensional temporal resolution (such as weeks or days) of spatial and temporal random effects could become feasible by using frequentist inference techniques for estimating fixed effects and random effect hyperparameters \citep{Waagepetersen.Guan.2009}, followed by INLA-based prediction of the intensity function for moderately large temporal subwindows of the observation and prediction period. Indeed, the Bayesian approach implemented in this paper is challenging but rewarding since complex fixed and random effects and associated uncertainties are estimated simultaneously in a very precise manner, but certain compromises with respect to the choice of inferential tools may be necessary when going towards even higher dimensions of data and model components.
Frequentist inference is used in our approach during a preprocessing step for interpolating meteorological data by kriging. In theory, joint Bayesian estimation of the meteorological model and the fire occurrence model could be possible with INLA, but in practice our two-step solution of mixing frequentist and Bayesian techniques is a pragmatic choice. Indeed, joint Bayesian estimation seems out of reach here due to the size of the dataset and the complexity of the latent Gaussian models. Morevover, to make INLA-based inference possible for modeling wildfire occurrences, we already had to devise a subsampling procedure for computational reasons. For reference, we mention \cite{Gomez-Rubio2020} who compares the uncertainties of estimates obtained by kriging and the SPDE-INLA approach on the \texttt{meuse} dataset in the R-package \texttt{gstat}, and finds better results for the Bayesian approach. However, this result may not have general validity, and in our case the nonseparable covariance structure of our kriging model is not implemented in R-INLA. Therefore, we think that using a frequentist kriging approach is the most pragmatic and reasonable choice here with respect to model flexibility, computational requirements, numerical stability of the estimation procedure, and quality of space-time prediction. The interpolation errors in kriging propagate through the model, but this is also the case for other sources of noise such as the positional uncertainty of wildfires, or the aggregation of land cover covariates to the DFCI grid. Our Bayesian modeling approach is designed to capture such remaining uncertainties due to the observation process in a sound, best possible way. Therefore, in this article, we do not pursue the goal to explicitly assess the propagation of kriging uncertainty in our model.

Moreover, we could combine our intensity model with a regression model for the size of wildfires (e.g., using the burnt surface as a mark of points) and focus on large wildfires in particular, such as by using INLA-based techniques for threshold exceedances \citep{Opitz.al.2018}. 
Finally, we point out that the use of static land use data and of wildfire ignition locations known only at a $2km$ resolution may have introduced some slight biases in the effects  estimated through our models. With the increasing availability of high-resolution high-frequency monitoring tools for land use and wildfire activity, such biases may be eliminated in future models. However, such increasingly massive datasets also call for new methodological developments at the interface of spatio-temporal statistical modeling,  which provides a sound basis for uncertainty assessment in the context of rare events,  and of powerful data mining and learning tools for ``big" data. 

\section*{References}

\bibliographystyle{apalike}
\bibliography{mybibfile}

\end{document}